\documentclass[apj]{emulateapj}
\shorttitle{Optical Color Variability of Nearby AGNs}
\shortauthors{Sakata et al.}

\begin{document}

\title{Long-Term Optical Continuum Color Variability of Nearby Active Galactic Nuclei}
\author{Yu Sakata\altaffilmark{1,2}, Takeo Minezaki\altaffilmark{1}, Yuzuru Yoshii\altaffilmark{1,3}, Yukiyasu Kobayashi\altaffilmark{4},\\ Shintaro Koshida\altaffilmark{2,4},
 Tsutomu Aoki\altaffilmark{5}, Keigo Enya\altaffilmark{6}, Hiroyuki Tomita\altaffilmark{5},\\ Masahiro Suganuma\altaffilmark{6}, Yuka Katsuno Uchimoto\altaffilmark{1}, Shota Sugawara\altaffilmark{1,2}} 

\altaffiltext{1}{Institute of Astronomy, School of Science, University of Tokyo, 2-21-1 Osawa, Mitaka, Tokyo 181-0015, Japan; yusakata@ioa.s.u-tokyo.ac.jp.}
\altaffiltext{2}{Department of Astronomy, School of Science, University of Tokyo, 7-3-1 Hongo, Bunkyo-ku, Tokyo 113-0013, Japan.}
\altaffiltext{3}{Research Center for the Early Universe, School of Science, University of Tokyo, 7-3-1 Hongo, Bunkyo-ku, Tokyo 113-0033, Japan.}
\altaffiltext{4}{National Astronomical Observatory, 2-21-1 Osawa, Mitaka, Tokyo 181-8588, Japan.}
\altaffiltext{5}{Kiso Observatory, Institute of Astronomy, School of Science, University of Tokyo, 10762-30 Mitake, Kiso, Nagano 397-0101, Japan.}
\altaffiltext{6}{Institute of Space and Astronomical Science, Japan Aerospace Exploration Agency, 3-1-1 Yoshinodai, Sagamihara, Kanagawa 229-8510, Japan.}

\begin{abstract}
	
We examine whether the spectral energy distribution of optical continuum emission 
of active galactic nuclei (AGNs) changes during flux variation, based on accurate 
and frequent monitoring observations of 11 nearby Seyfert galaxies and QSOs carried 
out in the $B$, $V$, and $I$ bands for seven years by the MAGNUM telescope.  
The multi-epoch flux data in any two different bands obtained on the same night show 
a very tight linear flux to flux relationship for all target AGNs.  
The flux of the host galaxy within the photometric aperture is carefully estimated by 
surface brightness fitting to available high-resolution HST images and MAGNUM images.  
The flux of narrow emission lines in the photometric bands is also estimated from available 
spectroscopic data.  
We find that the non-variable component of the host galaxy plus narrow 
emission lines for all target AGNs is located on the fainter extension of the linear regression 
line of multi-epoch flux data in the flux to flux diagram.  
This result strongly indicates that the spectral shape of AGN continuum emission 
in the optical region ($\sim4400$--$7900 \AA$) does not systematically change 
during flux variation.  
The trend of spectral hardening that optical continuum emission becomes bluer 
as it becomes brighter, which has been reported by many studies, is therefore 
interpreted as the domination of the variable component of the nearly constant 
spectral shape of an AGN as it brightens over the non-variable component of 
the host galaxy plus narrow lines, which is usually redder than AGN continuum emission.

\end{abstract}

\section{Introduction}

The UV-optical variability of type 1 AGNs such as Seyfert 1 galaxies and QSOs was 
found at the beginning of AGN studies in the 1960s and has been used as a powerful tool 
for understanding the nature of AGN central engines.  
The mechanism of their variability is under active discussion, and
 proposed models cover a diverse range of mechanisms including the instability of 
an accretion disk (Rees 1984; Kawaguchi et al. 1998), X-ray reprocessing 
(Krolik et al. 1991), star collision (Courvoisier et al. 1996), multiple supernovae 
(Terlevich et al. 1992), and gravitational microlensing (Hawkins 1993).  
None of these models, however, successfully explained more than a few properties
 of UV-optical variability (Vanden Berk et al. 2004). 

In efforts to construct a successful model of variability, photometric and 
spectroscopic monitoring observations have shown that the amplitude of 
UV-optical variability correlates well with various quantities of AGN luminosity, 
central black hole mass, accretion rate, redshift, rest-frame wavelength, 
and so on (Vanden Berk et al. 2004; Wold et al. 2007; Wilhite et al. 2008).  
Among these, a key relation that directly reflects the nature of the central 
engine is a blueing trend, in which AGNs become bluer when they brighten.  
Giveon et al. (1999) monitored 42 PG quasars in the $BR$ bands for seven years 
and showed such a blueing trend for at least half the quasars in their sample.  
Webb \& Malkan (2000) monitored 23 AGNs in their special $blue$, $yellow$, and 
$red$ bands for three months and showed a larger amplitude of variability in the 
blue band than that in the red band on timescales of months, while they found 
no color dependence of the variability amplitude on shorter timescales.  
Vanden Berk et al. (2004) used the two-epoch multicolor data of about 
25,000 quasars from the Sloan Digital Sky Survey (SDSS) and found an anti-correlation 
of variability amplitude with rest-frame wavelength especially at $\lambda < 4000 \AA$ 
in the UV-optical region.

The blueing trend, although observationally established, is known to admit a dual 
interpretation.  Some authors claim a naive interpretation, so-called spectral 
hardening, such that the variable component becomes brighter and gets bluer 
(e.g., Wamsteker et al. 1990; Giveon et al. 1999; Webb \& Malkan 2000; 
Vanden Berk et al. 2004; Wilhite et al. 2005), while others claim a different
 interpretation such that the variable component of constant blue color becomes brighter and
 increasingly dominates over the non-variable component of red color (e.g., Choloniewski 1981; 
Winkler et al. 1992; Winkler 1997; Paltani \& Walter 1996; Suganuma et al. 2006).   

The naive interpretation of spectral hardening was indeed countered in a convincing way 
by Winkler (1997), who monitored 92 Seyfert 1 galaxies in the $UBVRI$ bands for three years 
and showed a nearly linear flux to flux relationship in any two different bands obtained 
for each AGN on the same night. 
Using accurate and frequent $BV$ monitoring data for a few Seyfert 1 galaxies 
by the MAGNUM telescope (Kobayashi et al. 1998a, 1998b; Yoshii 2002; 
Yoshii, Kobayashi, \& Minezaki 2003), the linear relationship was undoubtedly 
confirmed by Suganuma et al. (2006) and Tomita et al. (2006).  
Winkler (1997) interpreted the linear relationship as indicating the constant color or 
constant spectral shape of the variable component, as originally proposed by 
\citet{1981AcA....31..293C}, with the assumption of non-variable component of 
the host galaxy located exactly on the fainter extension of the straight line 
fitted to the flux data in the flux to flux diagram.

The effect of host galaxy on the optical color variability of AGNs has been discussed 
by many authors \citep{1994AstL...20..606D,1999MNRAS.306..637G,2000ApJ...540..652W,2002ApJ...564..624T}.
\citet{1994AstL...20..606D} examined the correlation between the fluxes in two different 
bands of $UBVR$ for 30 Seyfert 1 galaxies and found,
opposing to the Choloniewski's interpretation, 
too large a scatter of their data from the linear relationship
to be consistent with the estimated errors for their data,
and
the host-galaxy component not on the fainter extension of the best-fit straight line for
five out of their 30 targets.
However, \citet{1997MNRAS.292..273W} presented the much tighter linear relationship than
the graphs displayed by \citet{1994AstL...20..606D}, with the scatter consistent
with the errors for most of their 92 targets, hence they concluded that
the large scatter found by \citet{1994AstL...20..606D} was likely caused by their 
underestimate of the measurement errors. 
%In addition, we note that
%for only 5 out of 30 objects
%\citet{1994AstL...20..606D} firmly
%established that the host-galaxy component is not located
%along the line of linear regression.

\citet{1999MNRAS.306..637G} estimated the host-galaxy flux for ten 
targets in their sample from the HST WFPC2 images obtained by \citet{1997ApJ...479..642B},
but did not explicitly take into account its contribution in their discussion 
on the spectral hardening.
\citet{2000ApJ...540..652W} estimated the host-galaxy flux for more than 
two thirds of their sample, but did not examine its effect for each of individual targets.  
In fact, they constructed a
spectral
model consisting of typical AGN and galaxy spectra, 
and examined the amplitude of multicolor variations.
They assumed that the host-galaxy component contributes
one third of the total flux in the yellow band
allowing 0.5 mag flux variation for the AGN component,
as an extreme case for a large contribution of host galaxy,
and found that the blue to red amplitude ratio was
larger than that expected from the model for two out of their 6 targets.
\citet{2002ApJ...564..624T} also constructed a similar spectral model to examine 
the color variability of AGNs in the sample of \citet{1999MNRAS.306..637G}.

We carried out multiwavelength monitoring observations of many type 1 AGNs using the 
MAGNUM telescope with high photometric accuracy, and high-resolution imaging 
data obtained by HST are now available for some of the MAGNUM targets.
Therefore, using the available high-resolution HST imaging data and the moderate-resolution 
ground-based images obtained by the MAGNUM telescope, we measure the flux contributions 
of the host galaxy for these AGNs with improved accuracy. 
Using previous studies, we also measure the flux contribution of narrow emission lines, 
which contaminate the photometry and affect the color variability as well.  
Then, subtracting the combined contribution of the non-variable 
host galaxy plus narrow lines component from the optical flux of each target AGN, 
we intend in this study to settle a long-standing problem as to whether the 
spectral shape of AGN continuum emission in the optical region
would change or remain constant with 
changing the brightness.

In Section \ref{obsandred} we describe the characteristics of the target AGNs, 
observations, reduction, and photometry.  
In Sections \ref{nonvarcompestimate} and \ref{contNL} we estimate the 
flux contributions of host galaxy and narrow lines, respectively.  
In Section \ref{testcolorvarchap} we obtain the optical flux of the target AGNs 
and discuss their color variability based on the linear correlation
between the fluxes in any two different bands and the location of the non-variable 
component of host galaxy plus narrow lines in the flux to flux diagram.  
In Section \ref{effectDT} we estimate the flux contribution of a dust torus to 
the observed optical flux.  
In Section \ref{discusschap} we discuss our interpretation of optical color variability of AGNs.  
In Section \ref{conclusionchap} we summarize our results.

\section{Multicolor Light Curve of AGNs}
\label{obsandred}

We describe the procedures of observation, data reduction, and photometry only briefly.  
More details are given in Suganuma et al. (2006).

\subsection{Observations}

Our targets were selected from a sample of AGNs included in the MAGNUM program 
(Yoshii 2002; Yoshii, Kobayashi, \& Minezaki 2003), on the condition that 
high-resolution HST/ACS images were available and their host galaxy contamination 
could be accurately estimated.  
We used archival HST/ACS image data obtained by Bradley Peterson 
(Proposal ID: 9851, 10516).  
As a result, our targets consist of nine Seyfert 1 galaxies, one radio-quiet QSO, 
and one radio-loud QSO.  
All of them are nearby ($z<0.06$) and are relatively low in luminosity (M$_B>-22$).  
Their characteristics are listed in Table \ref{objbasicparm}.  

Monitoring observations were made using the multicolor imaging photometer 
(MIP) mounted on the MAGNUM telescope (Kobayashi et al. 1998a, 1998b).  
The MIP has a field of view of $1'.5\times1'.5$ and obtains images simultaneously 
in the optical ($UBVRI$) and the near-infrared ($JHK$) bands by splitting the incident 
beam into two different detectors, an SITe CCD ($1024\times1024$ pixels, 
0.277 arcsec pixel$^{-1}$) and an SBRC InSb array  ($256\times256$ pixels, 
0.346 arcsec pixel$^{-1}$). 

The observational parameters of the target AGNs are listed in Table \ref{objobsparm}.  
Our targets were observed mostly in the $UBVIJHK$ bands for seven years, 
but some of them were observed in fewer bands and for shorter periods.  
The typical sampling interval in the $V$ band was 3 to 30 days (median value).  
Observations were made most frequently for the $(V,K)$ band pair: 
the second most frequently observed pair is $(B,H)$, and the third is $(U,J)$ or $(I,J)$.  
This ordering was selected because of our aim at studying dust reverberation of 
AGNs as the first priority (Minezaki et al. 2004; Suganuma et al. 2006).  
The FWHMs of the point-spread function (PSF) during the observations were
 typically $1''.0$--$1''.5$ in the optical and $0''.8$--$1''.0$ in the near-infrared.  
Monitoring observations were performed with our own highly automated observing 
system (Kobayashi et al. 2003). 

\subsection{Reduction and Photometry}

\begin{figure}
\epsscale{.80}
\plotone{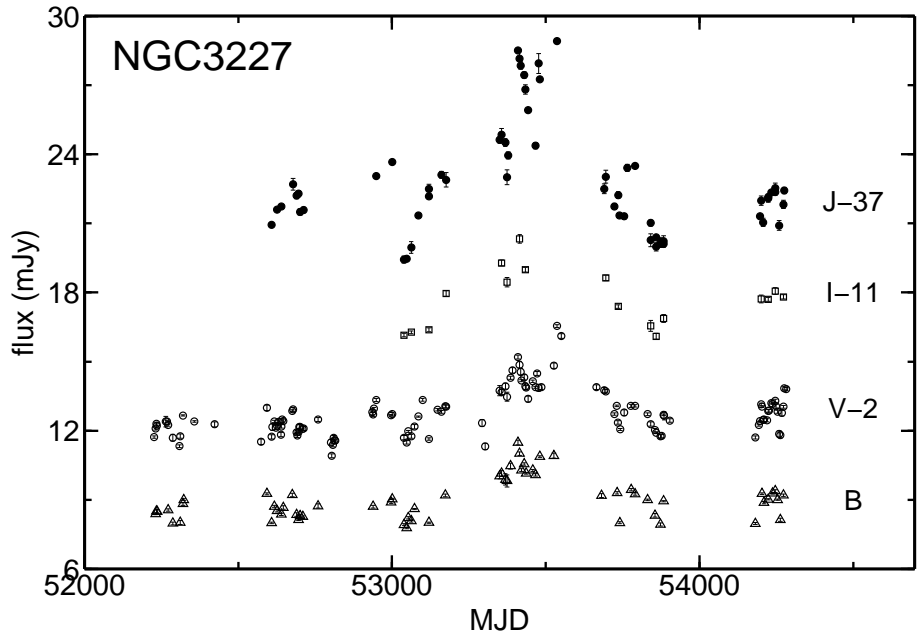}
\caption{Observed light curves in the $B$, $V$, $I$, and $J$ bands 
for the NGC3227 nucleus from 2001 November to 2007 July.\label{NGC3227lc}}
\end{figure}

\begin{figure}
\epsscale{.80}
\plotone{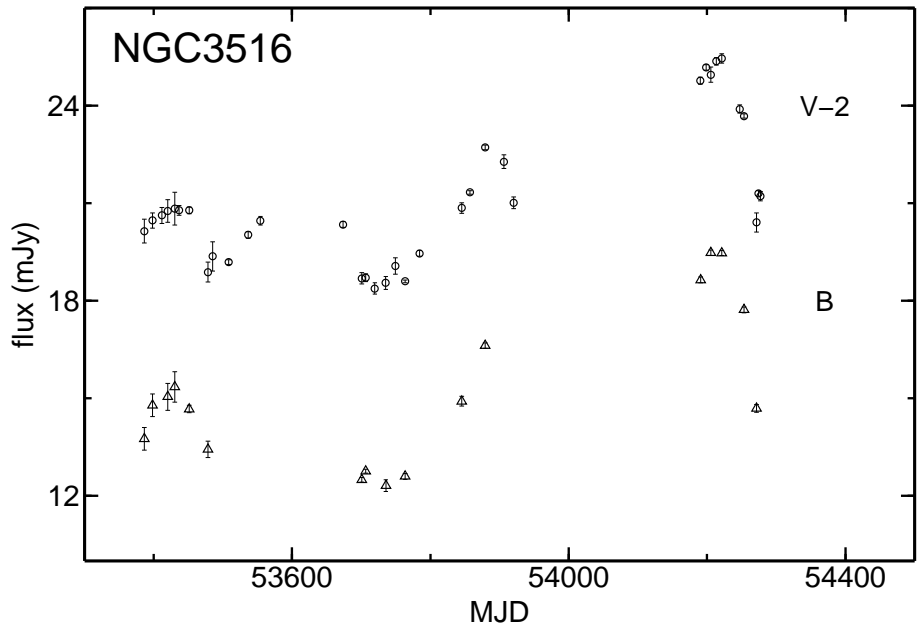}
\caption{Observed light curves in the $B$ and $V$ bands for the NGC3516 nucleus 
from 2005 January to 2007 June.\label{NGC3516lc}}
\end{figure}

\begin{figure}
\epsscale{.80}
\plotone{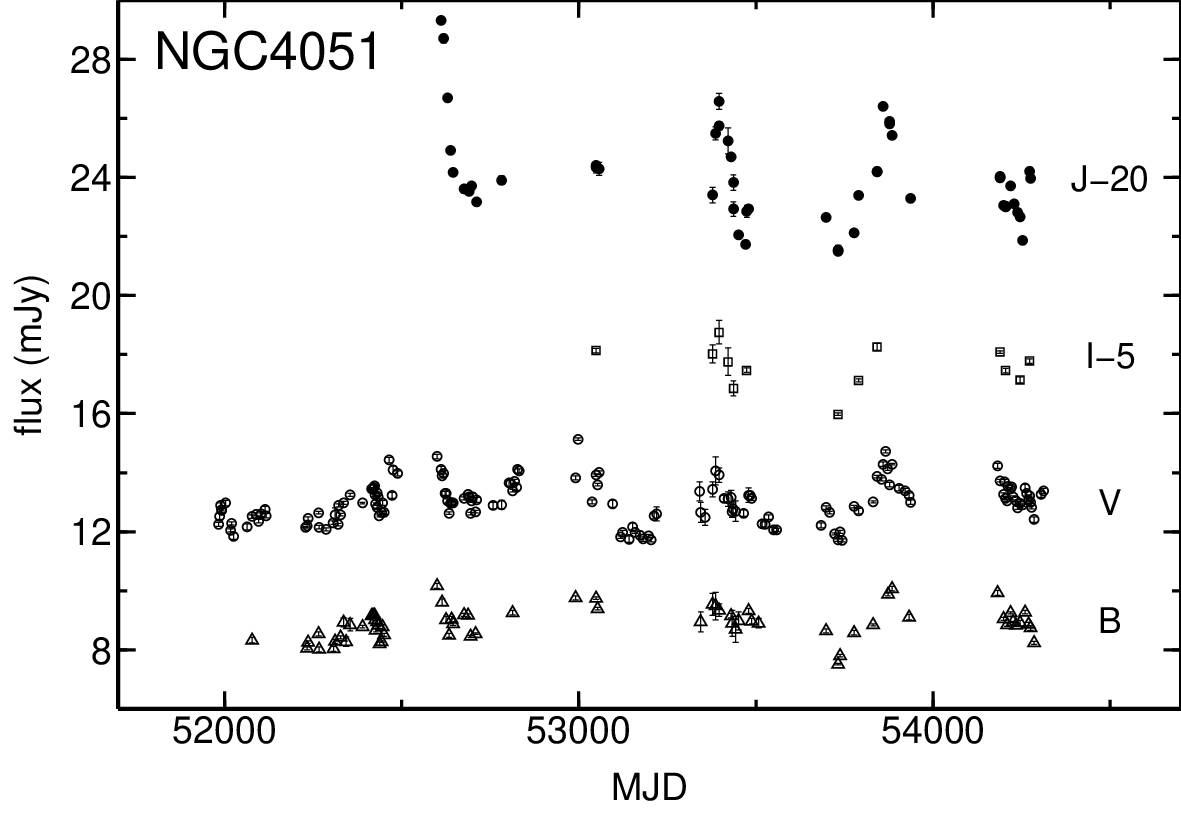}
\caption{Observed light curves in the $B$, $V$, $I$, and $J$ bands for 
the NGC4051 nucleus from 2001 March to 2007 July.\label{NGC4051lc}}
\end{figure}

\begin{figure}
\epsscale{.80}
\plotone{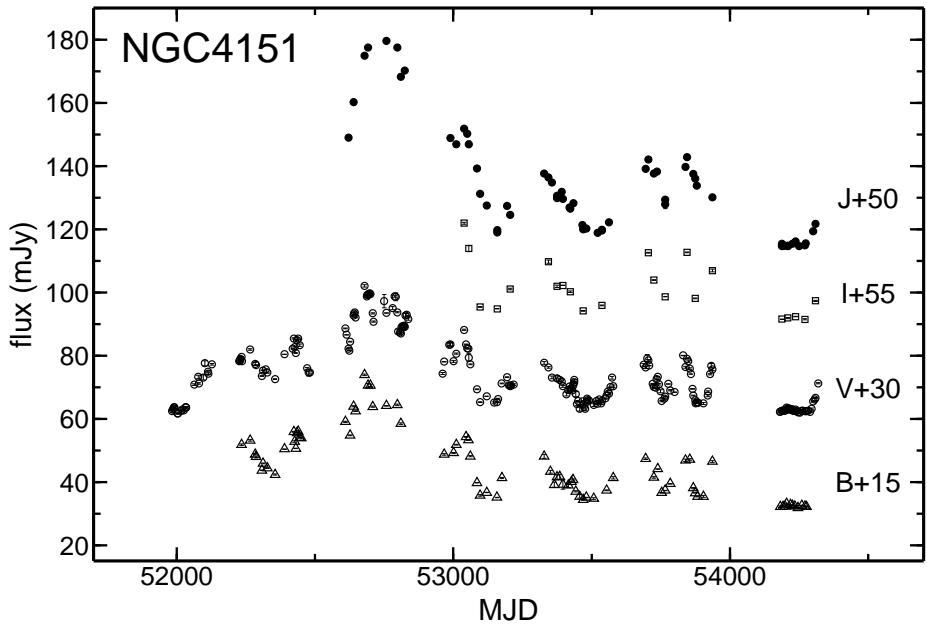}
\caption{Observed light curves in the $B$, $V$, $I$, and $J$ bands for 
the NGC4151 nucleus from 2001 March to 2007 August.\label{NGC4151lc}}
\end{figure}

\begin{figure}
\epsscale{.80}
\plotone{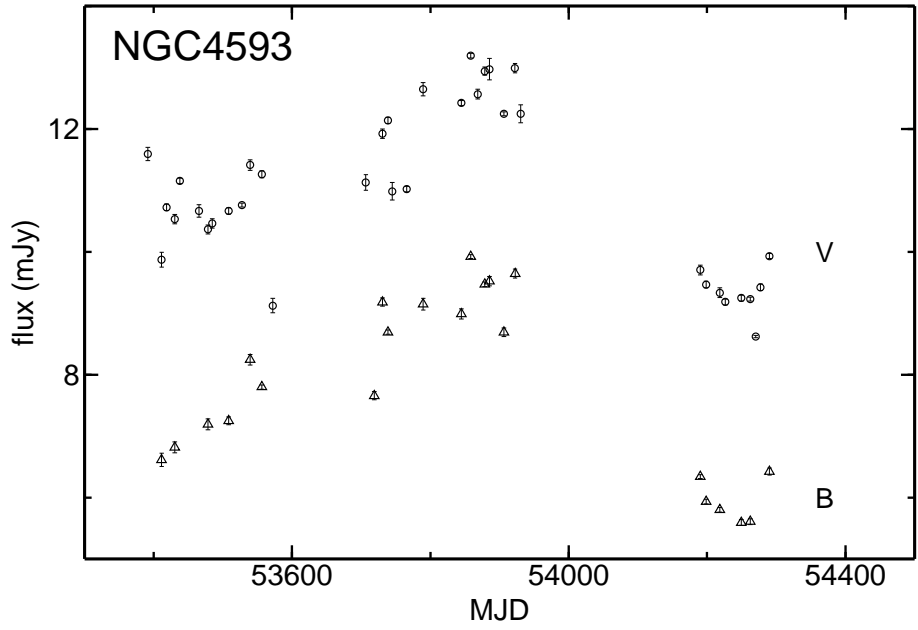}
\caption{Observed light curves in the $B$ and $V$ bands for the NGC4593 nucleus 
from 2005 January to 2007 July.\label{NGC4593lc}}
\end{figure}

\begin{figure}
\epsscale{.80}
\plotone{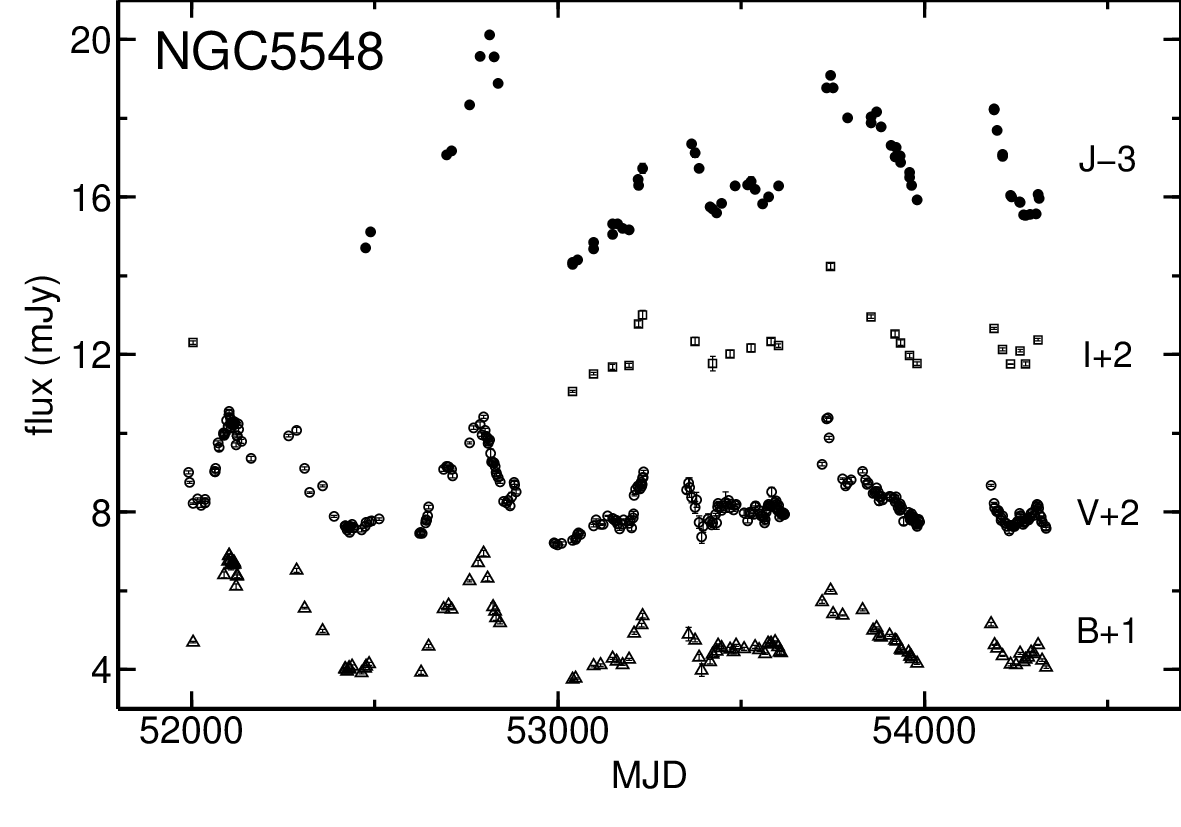}
\caption{Observed light curves in the $B$, $V$, $I$, and $J$ bands for the NGC5548 
nucleus from 2001 March to 2007 August.\label{NGC5548lc}}
\end{figure}

\begin{figure}
\epsscale{.80}
\plotone{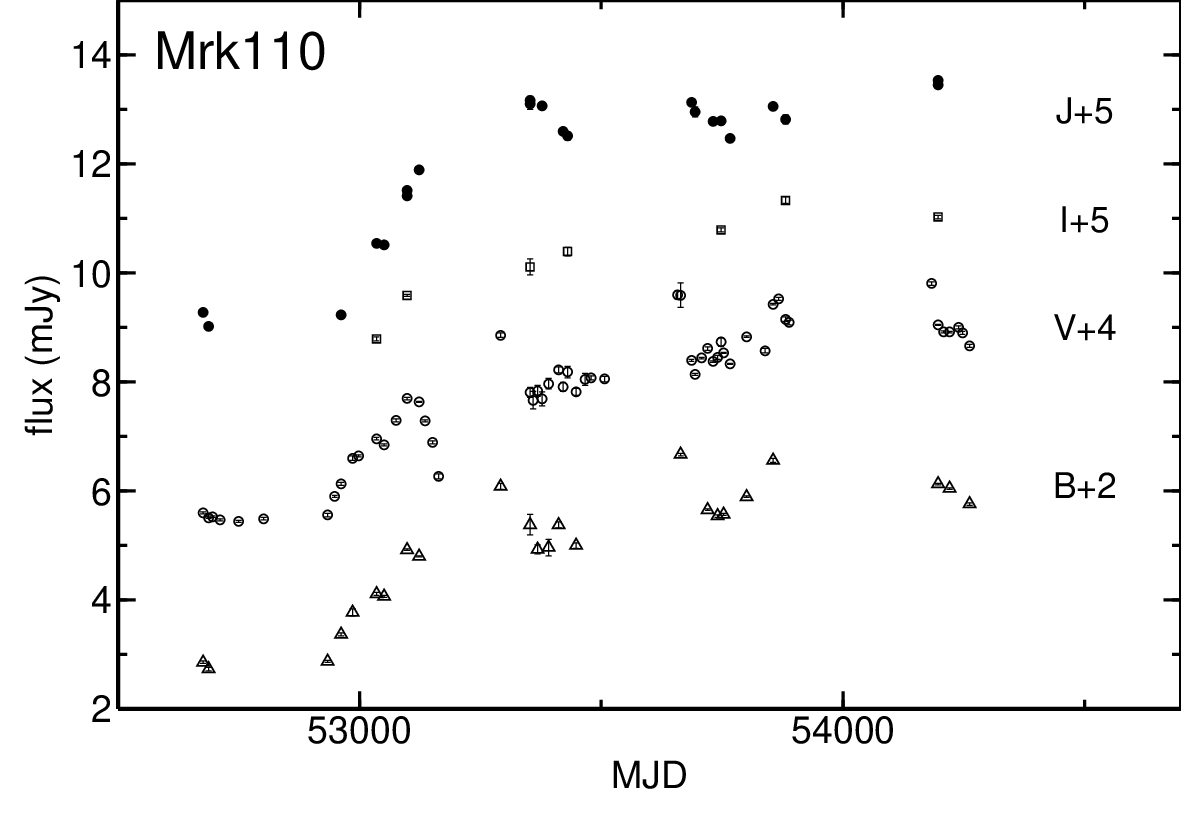}
\caption{Observed light curves in the $B$, $V$, $I$, and $J$ bands for 
the Mrk110 nucleus from 2003 February to 2007 August.\label{Mrk110lc}}
\end{figure}

\begin{figure}
\epsscale{.80}
\plotone{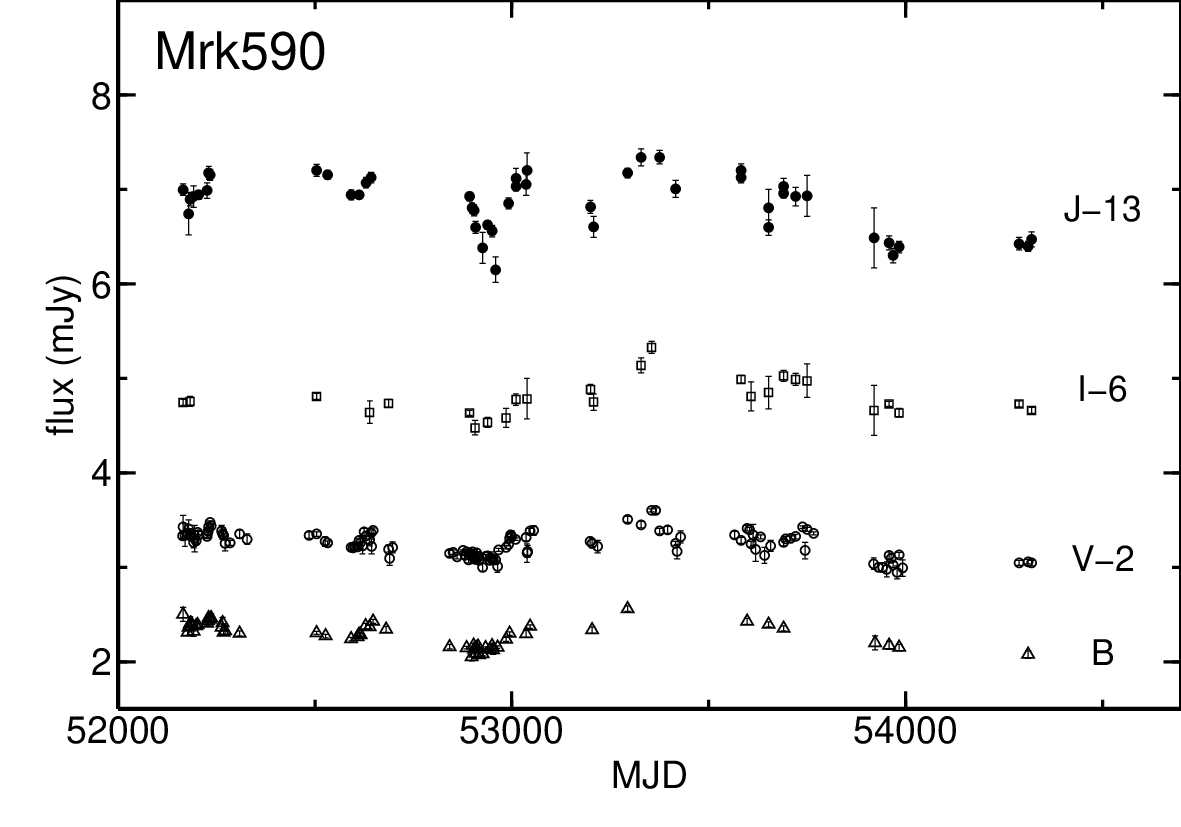}
\caption{Observed light curves in the $B$, $V$, $I$, and $J$ bands for 
the Mrk590 nucleus from 2001 September to 2007 August.\label{Mrk590lc}}
\end{figure}

\begin{figure}
\epsscale{.80}
\plotone{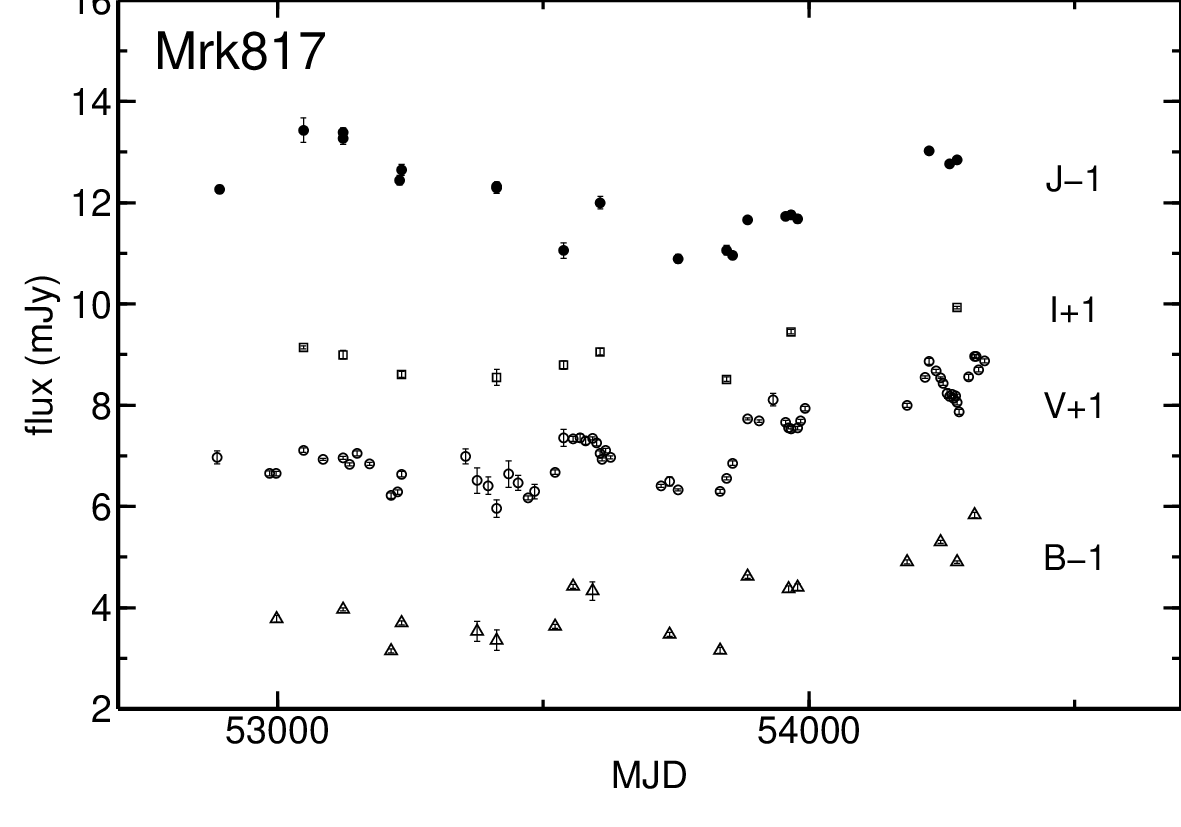}
\caption{Observed light curves in the $B$, $V$, $I$, and $J$ bands for 
the Mrk817 nucleus from 2003 September to 2007 August.\label{Mrk817lc}}
\end{figure}

\begin{figure}
\epsscale{.80}
\plotone{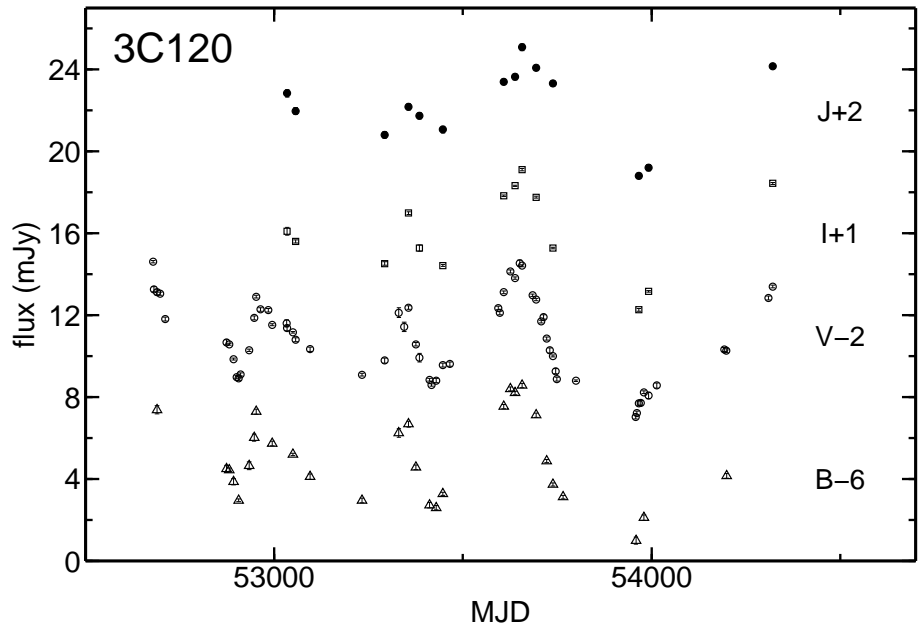}
\caption{Observed light curves in the $B$, $V$, $I$, and $J$ bands for 
the 3C120 nucleus from 2003 February to 2007 August.\label{3C120lc}}
\end{figure}

\begin{figure}
\epsscale{.80}
\plotone{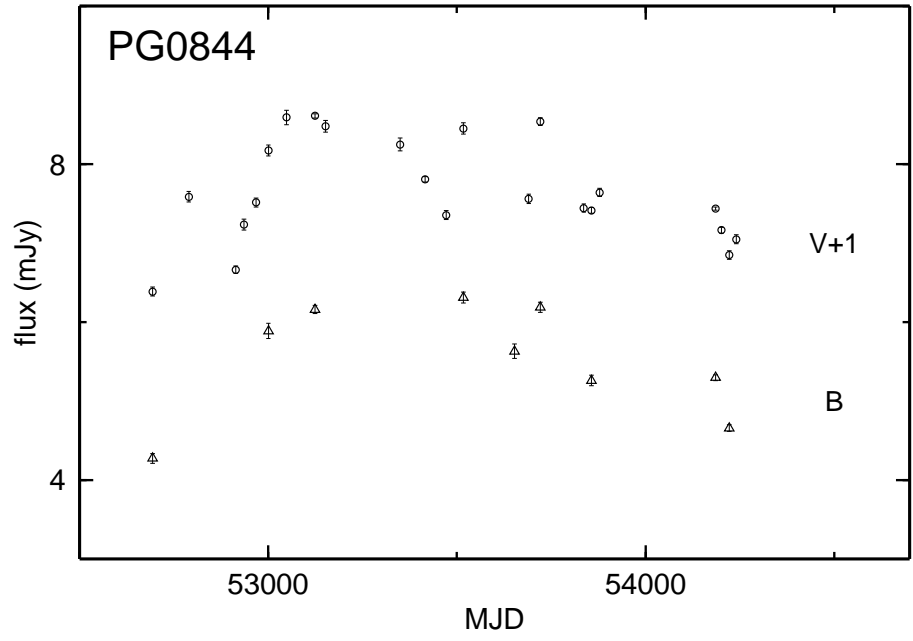}
\caption{Observed light curves in the $B$ and $V$ bands for 
the PG0844$+$349 nucleus from 2003 February to 2007 May.\label{PG0844lc}}
\end{figure}

The images were reduced using IRAF.  
We followed the standard procedures for image reduction,
 with small corrections for nonlinear detector response.  
The nucleus fluxes of our targets were
 measured relative to the nearby reference stars, and then the fluxes of 
the reference stars were calibrated.  
Aperture photometry within $\phi=8''.3$ with the sky reference of a 
$\phi=11''.1$--$13''.9$ annulus was applied to all images, 
and then the fluxes of the nuclei and reference stars were compared. 
We note that the sky annulus would include a significant portion of 
host-galaxy starlight for nearby target AGNs because of the apparently 
large extents of their host galaxies.
The fluxes of the reference stars were calibrated with respect to photometric 
standard stars taken from Landolt (1992) and Hunt et al. (1998).  
The uncertainty in the measurements was determined as follows: 
The rms variation of all measurements or the average photometric error for 
each image estimated from shot noise of photons, 
sky background noise, and readout noise, whichever was larger, was divided by 
square root of the number of images taken. 
Corrections for Galactic extinction were applied using the values given in 
the NASA/IPAC Extragalactic Database (NED; Schlegel, Finkbeiner, \& Davis 1998). 
Light curves based on aperture fluxes of all target AGNs are presented in 
Figures \ref{NGC3227lc} to \ref{PG0844lc}, where the offset fluxes 
of the host galaxy and narrow lines have not yet been subtracted. 

\section{Contribution of Host Galaxy}
\label{nonvarcompestimate}

In order to estimate the flux contribution of the host galaxy, 
we applied the surface brightness fitting using the two-dimensional image 
decomposition program GALFIT (Peng et al. 2002), which could fit the target image 
to analytic functions for the galaxy component of bulge and disk, 
plus a PSF for the nucleus and the sky contribution.  
Next, we applied aperture photometry with the same aperture size
and the same sky annulus
to a nucleus-free target image made by subtracting the best-fit PSF 
component from the target image.  
Since the surface brightness fitting technique would decompose the image into the 
galaxy component (in particular the bulge) and the nucleus component better when 
applied to higher spatial resolution images, we used HST/ACS high-resolution 
channel (HRC) data from the systematic AGN imaging survey by Bradley Peterson 
(Proposal ID: 9851, 10516).  
They were obtained in the F550M band. 
However, the flux contribution of the host galaxy in multiband 
is needed to be estimated to examine the color 
variation of the AGN optical continuum during flux variation. 
Thus, we applied the surface brightness fitting to the multiband MAGNUM images, 
fixing the host-galaxy shape parameters to 
those that were derived from the HST image, and applied aperture photometry 
to the nucleus-free MAGNUM images.  
In this way, we expect to improve the accuracy of host-galaxy estimation 
in ground-based images of lower spatial resolution.  

\subsection{Decomposition of HST/ACS/HRC Images into AGN and Host Galaxy}
\label{decomposeHST}

Although \citet{2006ApJ...644..133B, 2009ApJ...697..160B} applied the surface brightness 
fitting to the HST imaging data to decompose them into two components of AGN 
and host galaxy, we here perform such fitting independently and 
compare both results in Section \ref{checkHOST}.  
The archival HST data consist of three images with different exposure times 
(120 s, 300 s, 600 s) for each object.  
Since some of them were saturated,  we used the unsaturated image with 
maximum exposure time for each object.\footnote{
We also applied the surface brightness fitting to the HST imaging
data with three different exposure times by stacking them
following \citet{2006ApJ...644..133B},  and found little difference in 
the model parameters determined for the decomposed components.
In addition, the resulting host-galaxy fluxes using our decomposition 
parameters and those using the parameters derived by \citet{2009ApJ...697..160B}
are found to be consistent with each other,
as will be described in Section \ref{checkHOST}. }  
The image identifications for target AGNs are listed in Table \ref{BPMdata}.  
The PSF image was generated using the Tiny Tim package (Krist 1993), which could model 
the HST optics plus the camera and filter system specifications.  
The input parameters of Tiny Tim, the location on HRC and the spectral shape of 
the incident beam, were set to the nucleus location and the power-law spectrum of 
$f_{\nu}\propto \nu^0$  
\citep{2005ApJ...633..638W,tomita05,2006ApJ...652L..13T,2006ApJ...639...46S}, respectively.   

The HST images were corrected for the geometric distortion using the IRAF task of PyDrizzle.
First, the surface brightness fitting was applied to the distortion-uncorrected images 
with the distortion-uncorrected PSF image generated by Tiny Tim and the 
analytic functions for bulge and disk of galaxy model. 
This procedure significantly decreased
the fitting residual around the nucleus,
compared to the fitting applied to the distortion-corrected HST
image with the distortion-corrected PSF image.
Then, the correction for the distortion was applied to the best-fit model-galaxy image.  
Finally, surface brightness fitting was applied to the distortion-corrected model-galaxy 
image to derive the best-fit parameters of galaxy model (the fitting residuals here were 
found to be extremely small).  

\begin{figure}
\epsscale{.80}
\plotone{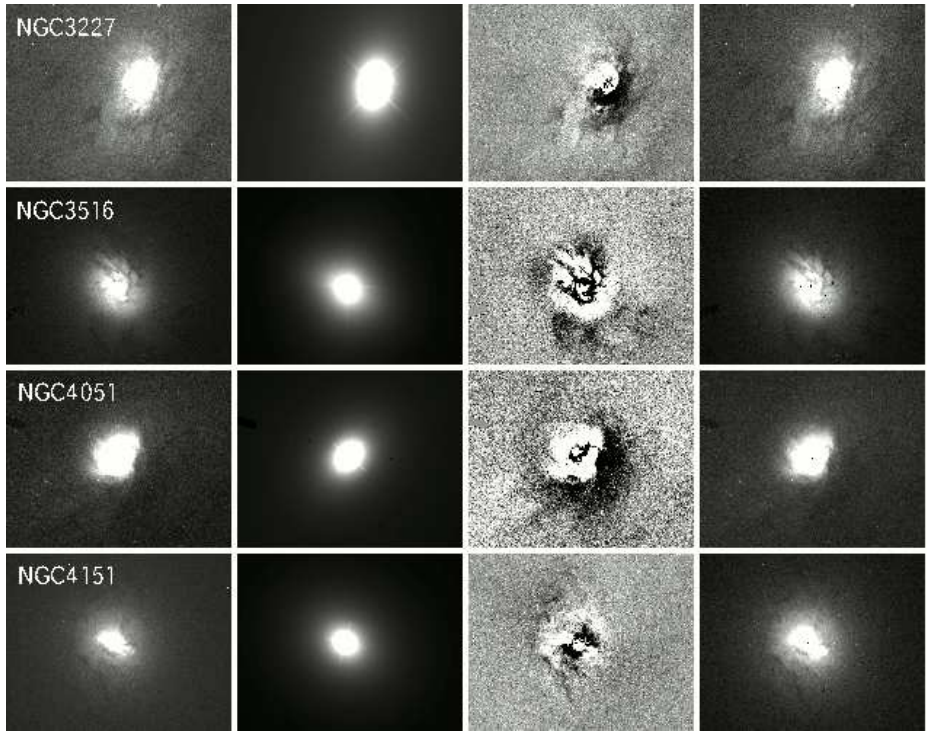}
\caption{The result of surface brightness fitting for the HST images of 
NGC3227, NGC3516, NGC4051 and NGC4151.  
Shown are the HST/ACS images corrected for the geometric distortion 
({\it left panels}), the model images of host galaxy$+$PSF ({\it second left panels}),
the residual images calculated by subtracting the host-galaxy model
from the nucleus-free images ({\it second right panels}), and the nucleus-free 
images calculated by subtracting the distortion-uncorrected model PSF
from the original HST/ACS images then correcting them for the geometric distortion
({\it right panels}).
All images are aligned as the north is up and the east is left.
\label{galfitfigurengc4051-3516_HST}}
\end{figure}

\begin{figure}
\epsscale{.80}
\plotone{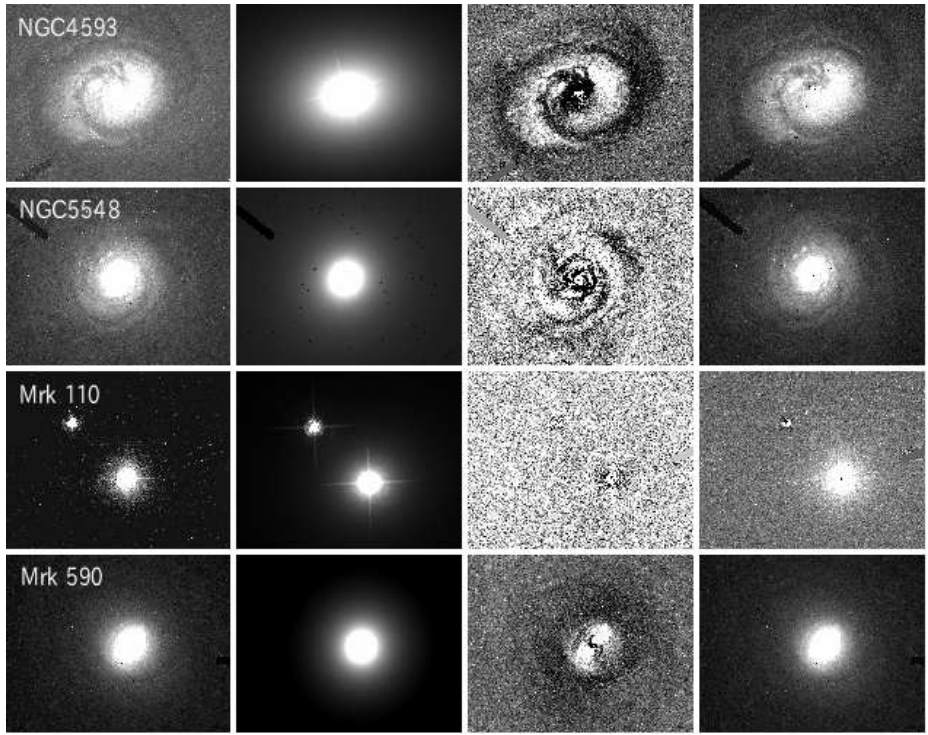}
\caption{The result of surface brightness fitting for the HST images of NGC4593, 
NGC5548, Mrk110 and Mrk590.  
Others are same as in Figure \ref{galfitfigurengc4051-3516_HST}.
\label{galfitfigurengc4593-817_HST}}
\end{figure}

\begin{figure}
\epsscale{.80}
\plotone{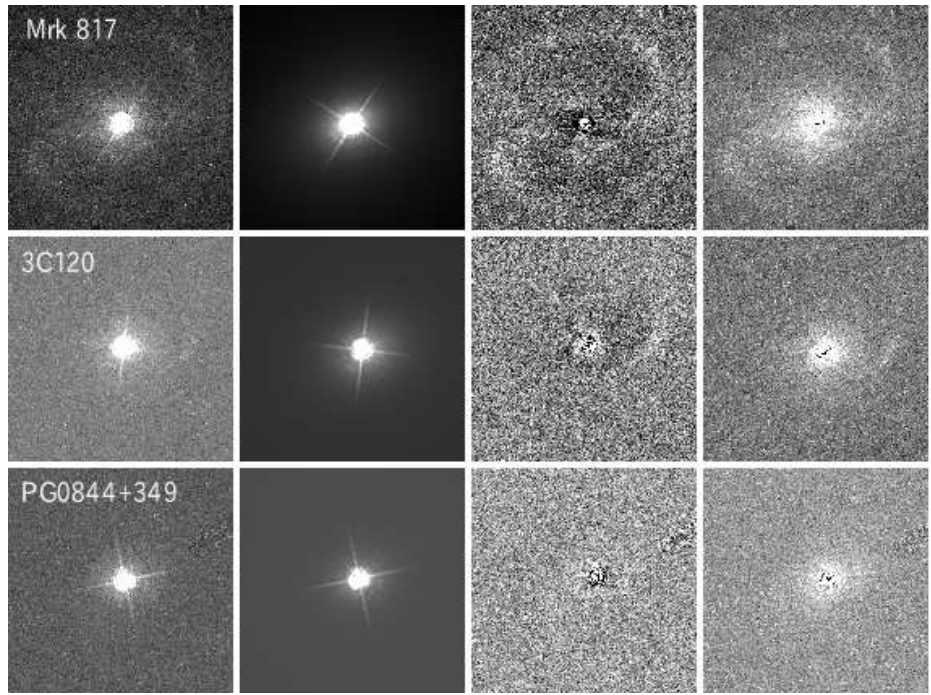}
\caption{The result of surface brightness fitting for the HST images of Mrk817, 
3C120 and PG0844+349.  Others are same as in Figure 
\ref{galfitfigurengc4051-3516_HST}.
\label{galfitfigurengc3C120-0844_HST}}
\end{figure}

The galaxy component was decomposed into bulge and disk fitted to 
a de Vaucouleurs' (1948) $R^{1/4}$ profile and an exponential profile,
respectively,
while only the de Vaucouleurs' profile 
was used for Mrk 110, Mrk 590 and PG0844$+$349,
because the decomposition into bulge and disk did not give a convergent result 
with any initial values of the parameters.
We performed the galaxy decomposition
by setting almost all parameters to be fitted 
except for the disk parameters;
Since
the disk parameters often converged 
at unrealistic values or diverged
without any restriction,
we fixed the disk center at the nucleus location 
and constrained the disk scale length and axis ratio within the
respective ranges
that were defined from the major and minor
isophotal diameters
of the target galaxy taken from NED. 
However, we adopted the best-fit disk scale length to be 
much smaller than those expected from the isophotal diameter
for NGC 3516, NGC 4593, and 3C120,
because small-scale structure in the central region
with complicated profile
seemed to be practically fitted by an exponential profile
for NGC 3516 and NGC 4593
\citep{1998ApJ...509..646F,2006ApJ...642..765K},
and because the isophotal diameter of a bulge-dominated galaxy
would not be a good indicator for the disk scale length
for 3C120.
Initial values of the parameters were set at those roughly estimated with IRAF, 
and then slightly different values were selected to examine how final values of 
the parameters varied with different initial values.  
The sky level of the parameters was fitted with its initial value set to be zero.  
When the fitted values were unrealistic,  
the best-fit values were searched for by changing the sky level
manually,
until the resultant sky fluxes were found to be comparable 
with those calculated from the HRC user's manual. 
Finally, we adopted the values of parameters that minimized the
reduced $\chi^2$.  
Figures \ref{galfitfigurengc4051-3516_HST}
to \ref{galfitfigurengc3C120-0844_HST} show 
the best-fit and residual images of the HST data.  
The best-fit parameters of the galaxy components are listed in Table
\ref{galfitparmresult}.

\subsection{Decomposition of MAGNUM Images into AGN and Host Galaxy}
\label{decomposeMAG}

To accurately estimate the host galaxy, we selected a moderate number of observations 
for each target and filter from MAGNUM monitoring data taken under good and stable 
seeing conditions and in the AGN-faint phase.  
The PSF image was made from the images of nearby reference stars for each target AGN.  
These stellar images were not taken simultaneously 
with the AGN images, so stable seeing was required for accurate fitting.  
In this way, more than five nights of images were selected in each band, although for some AGNs 
fewer than five nights were selected in each band because of the lack of appropriate data.  
Then, we combined the AGN images and stellar images from each night, respectively, to increase 
the signal to noise ratio and clipped the field region that is the same as the HST field for each AGN. 

\begin{figure}
\epsscale{.80}
\plotone{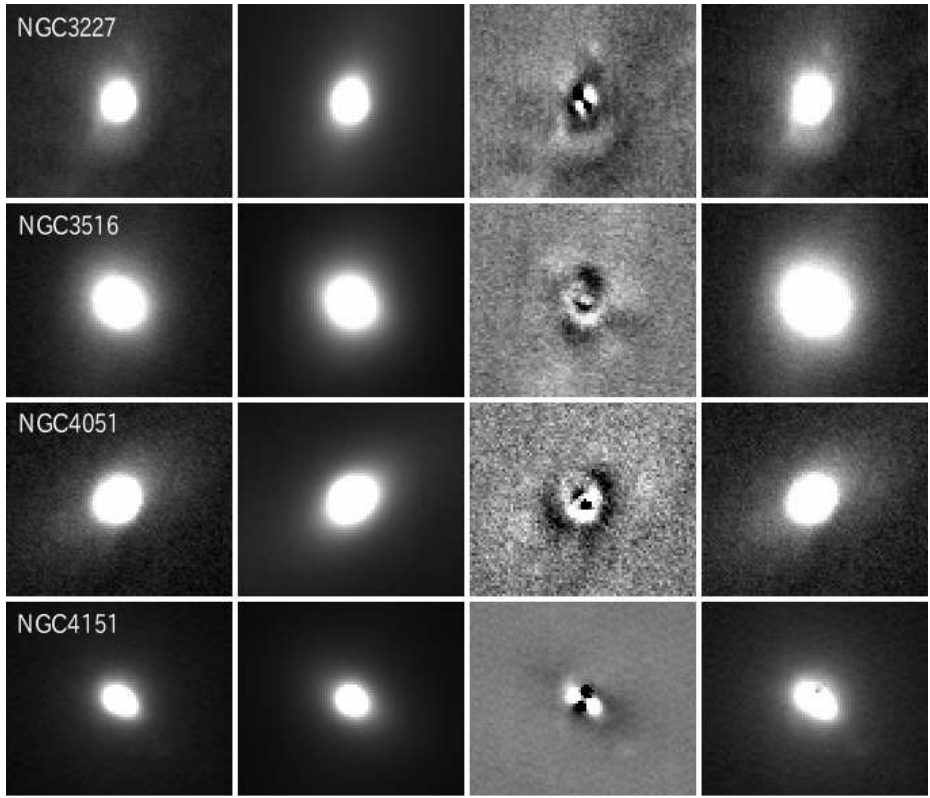}
\caption{The result of surface brightness fitting for the MAGNUM V-band images 
of NGC3227, NGC3516, NGC4051 and NGC4151. 
Shown are the original MAGNUM V-band images ({\it left panels}), 
the model images of host galaxy$+$PSF ({\it second left panels}),
the residual images calculated by subtracting the host-galaxy model
from the nucleus-free images ({\it second right panels}), and the nucleus-free 
images calculated by subtracting the model PSF
from the original images ({\it right panels}).
All images are aligned as the north is up and the east is left.
\label{galfitfigurengc4051-3516_MAG}}
\end{figure}

\begin{figure}
\epsscale{.80}
\plotone{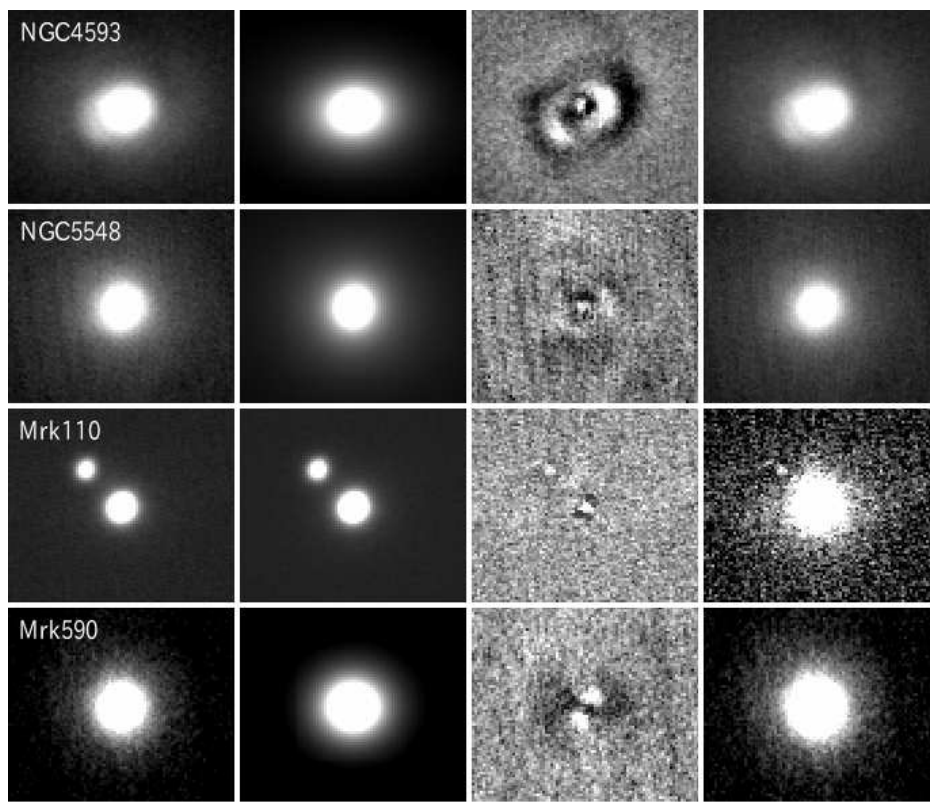}
\caption{The result of surface brightness fitting for the MAGNUM V-band images 
of NGC4593, NGC5548, Mrk110 and Mrk590.  
Others are same as in Figure \ref{galfitfigurengc4051-3516_MAG}.
\label{galfitfigurengc4593-817_MAG}}
\end{figure}

\begin{figure}
\epsscale{.80}
\plotone{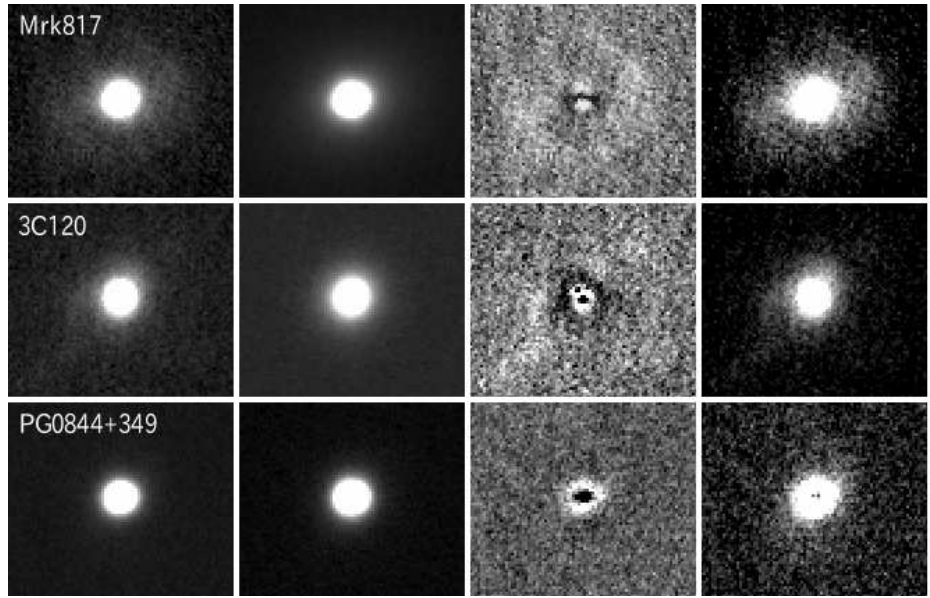}
\caption{The result of surface brightness fitting for the MAGNUM V-band images 
of Mrk817, 3C120 and PG0844+349.  
Others are same as in Figure \ref{galfitfigurengc4051-3516_MAG}.
\label{galfitfigurengc3C120-0844_MAG}}
\end{figure}

These combined and clipped images of target AGNs for each night were fitted to the PSF 
made as described for the nucleus component and to analytic functions for the galaxy 
component of bulge and disk, with the parameters of host-galaxy shape  
(scale length and axis ratio) fixed at the best-fit values of the surface brightness fitting
 using the HST data. 
The sky level was fixed at the mean sky value of reference star images.  
When $\chi^2$ was obviously large, we did not fix the disk scale length.  
This occurred mainly in the $B$ and $I$ images,
probably because of possible radial color gradient in a galaxy, arising either 
from stellar population gradient or from metallicity gradient (e.g., de Jong 1996; 
Taylor et al. 2005).
However, this had eventually little effect in estimating the host galaxy flux 
(differences were less than a few percent).  
Finally, we applied the aperture photometry with the same aperture size to a nucleus-free image 
made by subtracting the best-fit PSF component from the original image.
With a moderate number of independent observations available, the host-galaxy flux 
was estimated as the average of those fluxes of all observations, 
and its error was estimated as their standard deviation.  
Figures \ref{galfitfigurengc4051-3516_MAG}
to \ref{galfitfigurengc3C120-0844_MAG} 
show the best-fit and residual images of the MAGNUM data.  
The estimated values of the host-galaxy flux for target AGNs 
within the monitoring aperture are listed in Table \ref{MAGNUMhostresult}.

\subsection{Consistency Check of Host Galaxy Estimates}
\label{checkHOST}

In order to improve the estimation of the host-galaxy flux in the ground-based image, 
we decomposed the MAGNUM image into the components of AGN and host galaxy using 
the shape parameters of host galaxy derived from the HST image.\footnote{Since we intended to estimate the host-galaxy flux within
the aperture around the center and not to study its whole structure, 
we did not fit the high-resolution 
HST image and the wide-area MAGNUM image together.} 
Here, we compare our results of the host-galaxy flux from the MAGNUM images in 
the V band with those estimated from the HST images in the F550M band,
because the effective wavelengths of these bands are very close to each other.  
We note that they are based on the aperture photometry 
with the same aperture size to the nucleus-free image made by subtracting 
the best-fit PSF component from the HST image. 
The procedure of image decomposition is described in Section \ref{decomposeHST}.
The error of the host-galaxy flux was estimated as follows:
An artificial image was generated with the best-fit parameters
by adding artificial noise.
Then, the PSF flux of the artificial image was
increased and decreased with the total flux of PSF and bulge kept constant,
and the reduced $\chi ^2$ was calculated each time.
Finally, the error was estimated as the range of the host-galaxy flux of
the artificial image such that
the reduced $\chi ^2$ calculated above is the same as that of
the observed image fitted with the best-fit parameter values.

\begin{figure}

\plotone{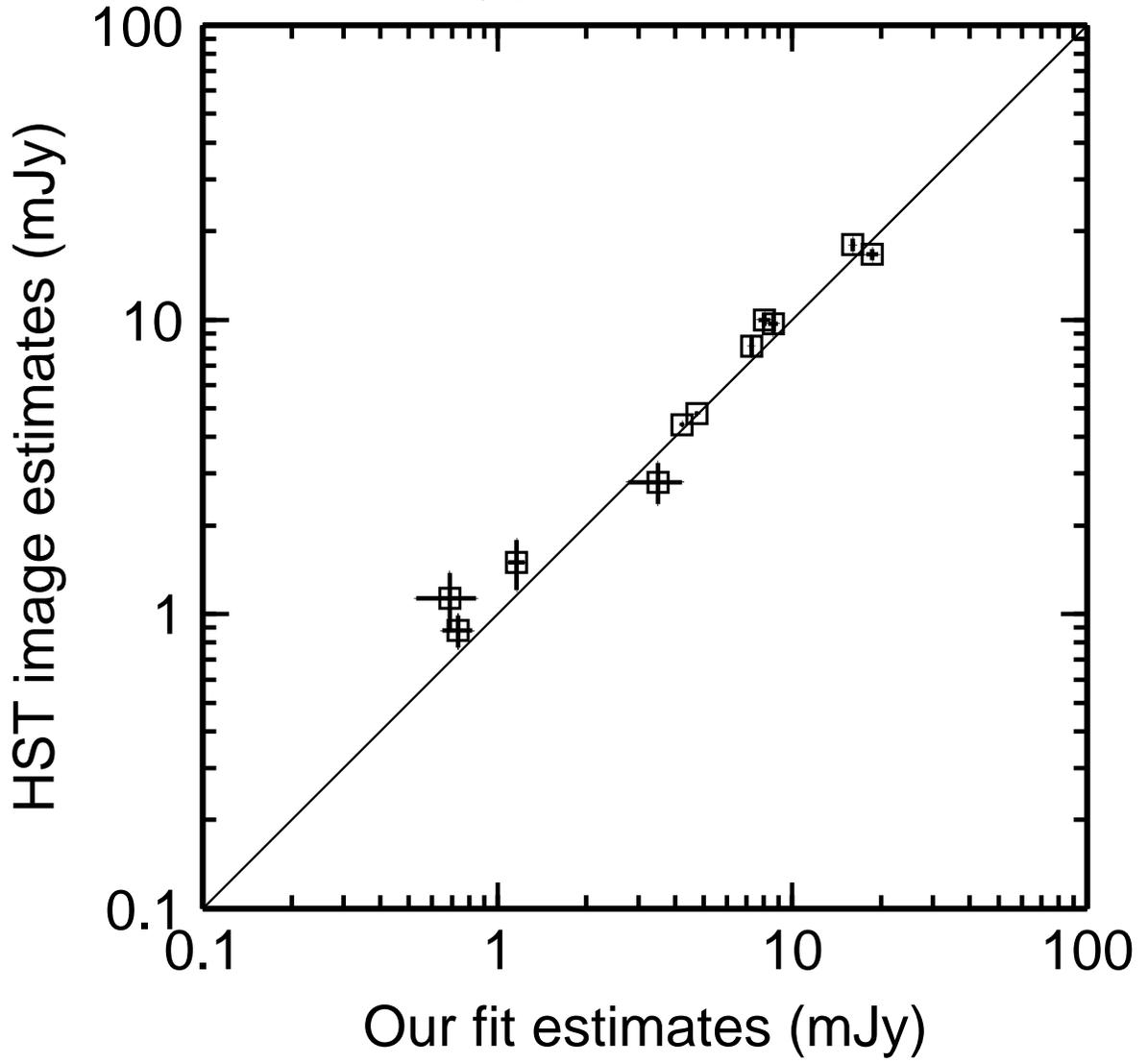}
\caption{The comparison between the host galaxy flux estimated from MAGNUM 
$V$-band images and that from HST F550M-band images, for our 11 targets. 
The diagonal line represents the line of agreement between the two estimates. 
\label{comparisonHST}}
\end{figure}

Figure \ref{comparisonHST} shows the host-galaxy fluxes of the 11 target AGNs 
in the V band for MAGNUM and in the F550M band for HST. 
We have found that both results are consistent with each 
other within a 3$\sigma $ level of errors.
We have also found that the systematic difference between them is as small as $1.4$ \% 
and the standard deviation is $18$ \%, which further decreases to $11$ \% 
when the eight targets with brighter host galaxies are selected. 
This good agreement suggests that our estimation of the host-galaxy flux 
in the MAGNUM image provides an accuracy comparable 
to that in the HST image.

\begin{figure}
\plotone{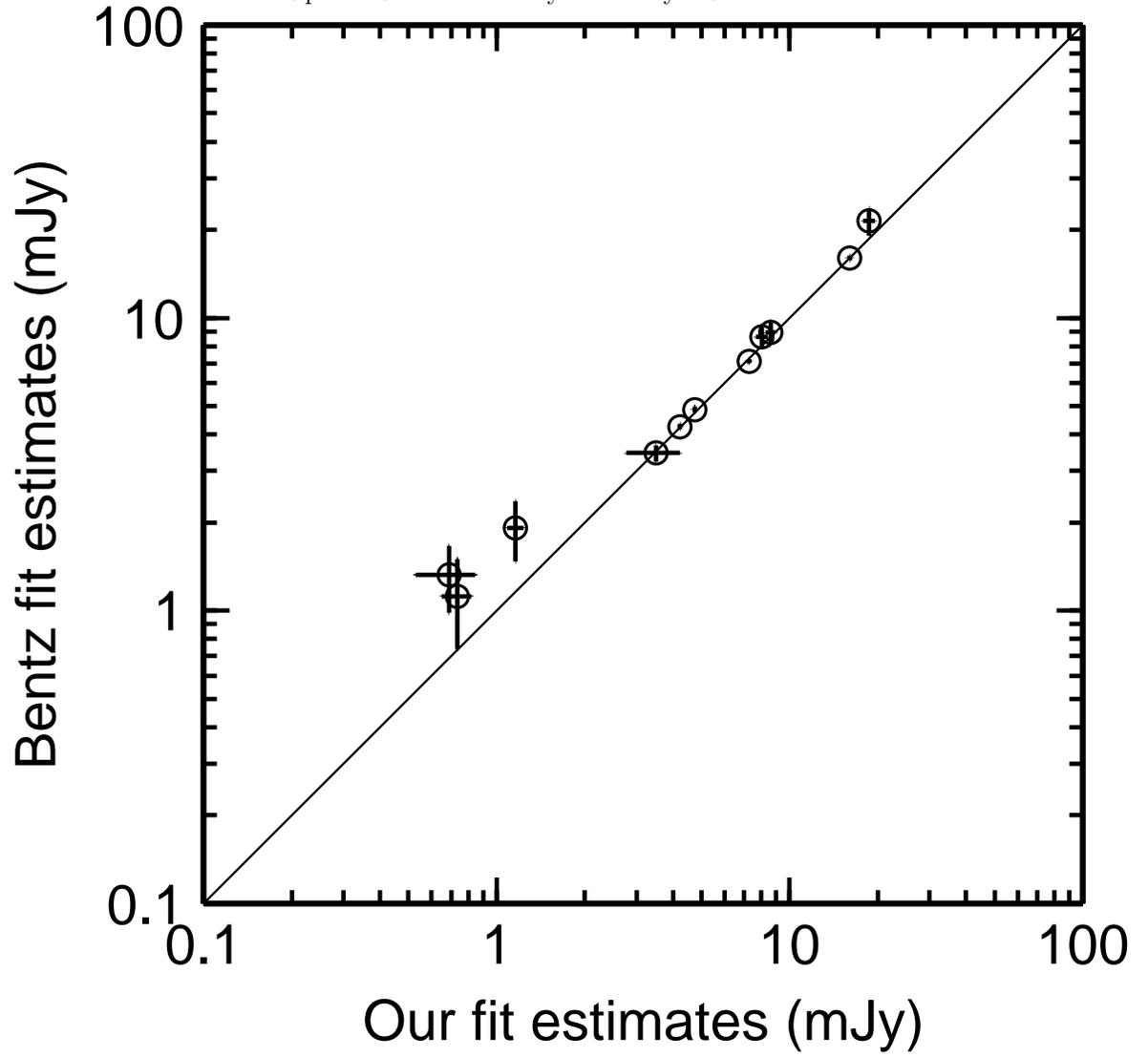}
\caption{The comparison between the host galaxy fluxes estimated from MAGNUM 
$V$-band images decomposed with the galaxy-shape parameters being fixed 
at ours and with that of Bentz et al. (2009). 
The diagonal line represents the line of agreement between the two estimates. 
\label{comparisonBentz}}
\end{figure}

Bentz et al. (2009) performed the surface brightness fitting of 
host galaxy component using the Sersic (1968) profile.  Since his 
method allows a more general distribution compared with our use 
of de Vaucouleurs' profile and an exponential profile, 
it is instructive to decompose the MAGNUM image in the same way 
as described in Section \ref{decomposeMAG}, 
but with the parameter values of host-galaxy shape 
in Table 4 in Bentz et al. (2009). 
Figure \ref{comparisonBentz} shows the host-galaxy fluxes of 
11 target AGNs with the parameter values of host-galaxy shape 
adopted by \citet{2009ApJ...697..160B}, 
in comparison of those with our own values of such parameters.  
We have found that the systematic difference is only $0.3$\%
and the standard deviation is $4.3$\% for the eight targets with brighter host galaxies,
while for other three targets with fainter host galaxies, 
both results are consistent with each other within a 2$\sigma$ level of errors. 
This good agreement between the host-galaxy fluxes estimated with two 
sets of host-galaxy parameters suggests that the uncertainty of AGN 
and galaxy decomposition in the HST images is not significant in our analysis below.

\subsection{Host Galaxy Color}

The estimated colors of host galaxies within the photometric aperture 
are listed in Table \ref{hostcolor}.  
For comparison, typical colors of galaxies (Fukugita et al. 1995) and bulge (Kinney et al. 1996)
 are also presented in Table \ref{hostcolor}.  
For nearby targets (NGC 4051, NGC 3227, NGC 3516, NGC 4593, NGC 5548, Mrk 590, Mrk 817), 
the dominant host-galaxy component within the aperture is identified with a bulge, 
and the $B-V$ and $V-I$ colors are comparable to those colors of bulge or slightly bluer.  
This result is interpreted as meaning that the host flux within the aperture is mainly from the 
bulge or slightly contaminated by the disk, which is generally bluer than the bulge.  
For NGC 4151, the $B-V$ color is interpreted as above, but the $V-I$ color is slightly bluer 
than the bulge.  
However, the small contamination of the extended $[$\ion{O}{3}$]$ emission into 
the host-galaxy fluxes in the $B$ and $V$ bands could explain 
the colors of NGC 4151's host galaxy.
Observation by Yoshida et al. (1993) show that the $[$\ion{O}{3}$]\lambda 4959,5007$ emission lines 
of NGC 4151 come not only from the nucleus but also from the extended region 
($\sim $10''$\times $10'' around the nucleus), and the flux of extended $[$\ion{O}{3}$]$ 
emission is about $10$--$20\%$ of nucleus $[$\ion{O}{3}$]$ emission.  
The $[$\ion{O}{3}$]$ emission that contributed to the $V$-band flux would make 
the $V-I$ color slightly bluer, while it happens to contribute to both the $B$-band and $V$-band 
fluxes similarly to the color of the host galaxy. 
Consequently, the $B-V$ color stays almost unchanged. 

For relatively distant targets (3C 120, Mrk 110, PG 0844+349), the flux of the host galaxy 
within the aperture contains nearly all the flux from the host galaxy. 
For Mrk 110, the dominant host-galaxy component within the aperture is identified 
as a late-type spiral disk, and the $B-V$ and $V-I$ colors are comparable to those of a typical 
spiral galaxy, but for 3C 120 and PG 0844+349, although these dominant host galaxies are 
identified as early-type galaxies, their color tends to be bluer than that of a typical early-type 
galaxy.  
However, many authors have reported that the host-galaxy colors of QSOs tend to be
 bluer than that of a typical early-type galaxy (Kauffmann et al. 2003; Sanchez et al. 2004; 
Schweitzer et al. 2006; Jahnke et al. 2004, 2007).  
Therefore, the host-galaxy colors we estimated are consistent with the trend reported 
by previous studies.

\section{Contribution of Narrow Lines}
\label{contNL}

Observations of nearby AGNs show that the spatial extent of the narrow-line region is typically 
around 100 pc or larger (Schmitt et al. 2003), so the narrow-line flux does not respond 
to variations in incident flux from the nucleus and is generally regarded as non-variable 
over 100 yr.  
Therefore, it is possible to estimate the narrow-line flux from past spectral data.  
We considered only strong narrow lines ([\ion{O}{3}]$\lambda$4959,5007, H$\beta$, 
H$\gamma$), because these lines occupy a large fraction of the optical narrow-line flux 
(about 90$\%$ or more, except for the small blue bump) 
and the flux of other lines is hardly estimated. 

Compilation of past [\ion{O}{3}]$\lambda$5007 observations by Whittle et al. (1992) 
indicates that the narrow-line fluxes of very nearby AGNs 
within different apertures are not always the same.  
The scatter of narrow-line fluxes in large apertures of $7''$--$15''$ is smaller 
than that in small aperture of $2''$--$4''$.  
They interpreted this difference as the seeing effect and angular size 
effect (a large aperture covers most of extended narrow line regions).  
For [\ion{O}{3}]$\lambda 5007$, we adopted the mean fluxes of large-aperture 
observations presented in Whittle et al. (1992), because the aperture size is 
comparable to our photometric aperture ($\phi=8''.3 $), and the scatter of 
narrow-line fluxes is small in many large-aperture observations. 
For Mrk 110, Mrk 590, and Mrk 817, we adopted the mean fluxes of small-aperture 
observations in Whittle et al. (1992) because large-aperture data were not available. 
We note that the quality of these data is poorer.  
[\ion{O}{3}]$\lambda 5007$ data for 3C 120 and PG 0844+349 were not 
presented in Whittle et al. (1992), so we adopted data from 
Tadhunter et al. (1993) obtained with a $5''$ slit and from Miller et al. (1992) 
obtained with a $6''$ slit, respectively.  

For $[$\ion{O}{3}$]\lambda$4959, we adopted fluxes computed with a theoretical line 
ratio of [\ion{O}{3}]$\lambda 5007/$[\ion{O}{3}]$\lambda4959 = 3.01$ 
(Storey \& Zeippen 2000).  
For H$\beta$, we adopted fluxes for some targets from 
Kaspi et al. (2005), and for others those computed with a roughly estimated line 
ratio of $[$\ion{O}{3}$]\lambda5007/\textrm{H}\beta \simeq 10$ 
(Veilleux \& Osterbrock 1987).  For $H\gamma$, we adopted fluxes computed with a 
theoretical line ratio of $\textrm{H}\gamma/\textrm{H}\beta \simeq 0.45$ 
(Case B: low-density limit and $T=5000$ K, Osterbrock \& Ferland 2005). 
All fluxes were corrected for Galactic extinction using the linearly-interpolated NED value.  
The adopted values of narrow-line flux are listed in Table \ref{NLfluxdata}. 

Finally, we estimated the flux contribution of narrow lines in the wavelength coverage of 
each filter by convolving the narrow-line spectrum with the MAGNUM filter transmission curve.  
We assumed a $\delta$-function for narrow-line profiles for simplicity, because the width 
of the narrow lines is so small that the filter transmission can be taken as constant 
at their wavelengths.  
The estimated values of flux contribution of the narrow lines in the $B$ and $V$ 
bands are listed in Table \ref{NLfinalresult}.  
We note that the narrow-line component has a minor contribution to the total flux 
in aperture, and the host galaxy component always makes a major contribution.

\section{Examination of Color Variability}
\label{testcolorvarchap}

In order to examine the color variability of optical continuum emission 
of type 1 AGNs with their flux variation, 
we applied flux to flux plot analysis, originally proposed by Cho{\l}oniewski (1981).  
Using the $UBV$ monitoring data of Seyfert galaxies, Cho{\l}oniewski (1981) plotted 
the data of each target obtained at different epochs on a diagram of the 
total flux in one band versus the total flux in another band, 
and first noticed a linear correlation between the fluxes of an AGN 
in two different optical bands.  
This has subsequently been confirmed by many authors 
(Winkler et al. 1992; Winkler 1997; Suganuma et al. 2006; Tomita et al. 2006), 
and it is assumed that the vector describing the total observed 
flux in the flux to flux diagram is the sum of a constant flux vector of 
starlight of the host galaxy and a variable flux vector of the AGN continuum 
that is constant in direction but not in magnitude.  
Assuming a variable flux vector of the AGN continuum that does not change its direction 
is equivalent to assuming the constant spectral shape of the AGN continuum. 

We first examine the linear correlation between the fluxes in two different bands by plotting 
our long-term monitoring data with high photometric accuracy in a flux to flux diagram.  
Then, by summing the host-galaxy flux and narrow-line contribution, 
we locate the non-variable component in the flux to flux diagram.  
In this way, we are able to verify the simple two-vector model, 
by examining whether the non-variable component is located on the fainter extension 
of the linear correlation line.

\subsection{Flux to Flux Plot and Linear Fit}

\begin{figure}
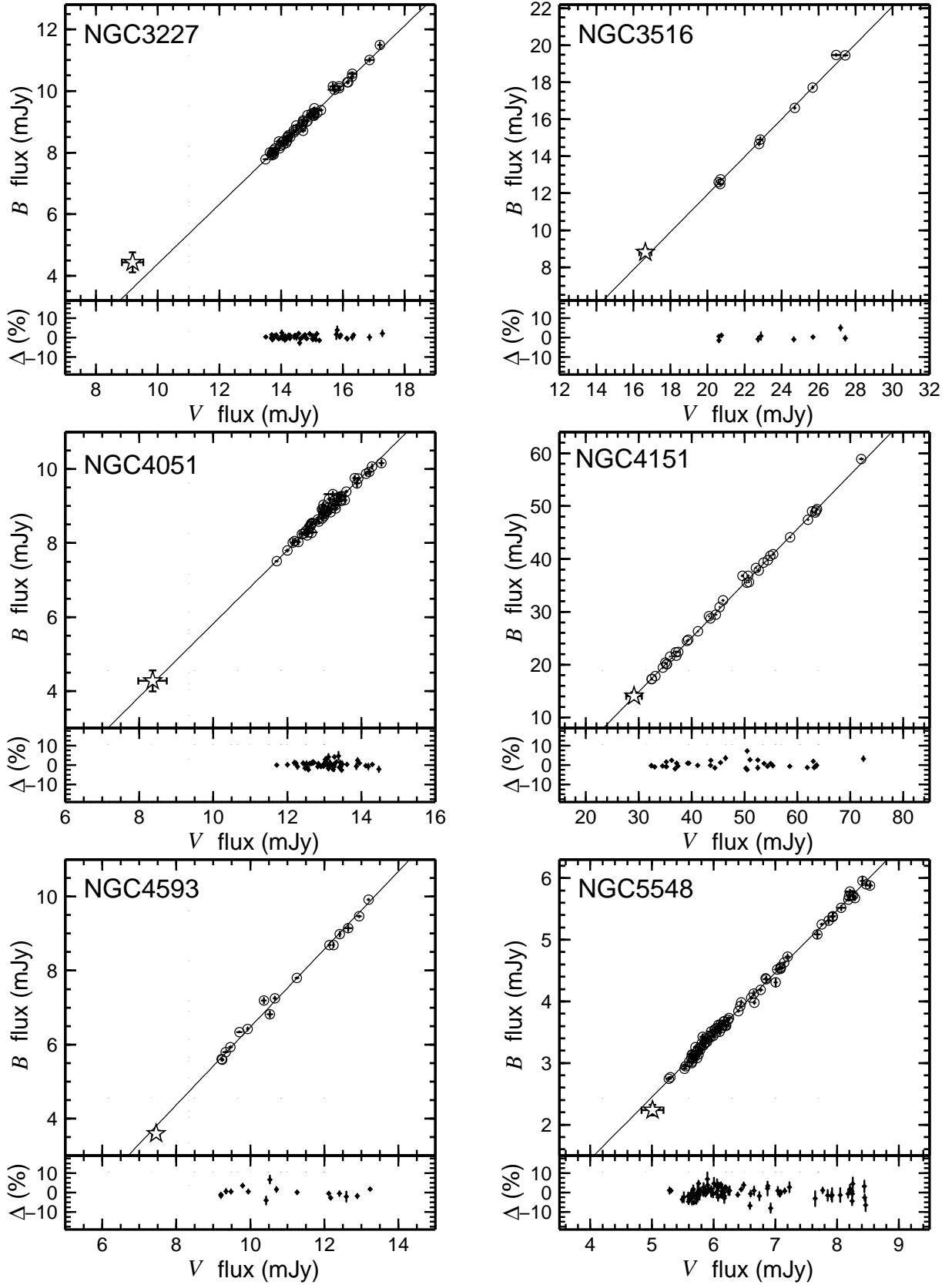


\plottwo{f20a.eps}{f20b.eps}

\plottwo{f20c.eps}{f20d.eps}

\plottwo{f20e.eps}{f20f.eps}

\caption{The $B$-band flux to $V$-band flux diagrams for target AGNs 
of NGC 3227, NGC 3516, NGC 4051, NGC 4151, NGC 4593 and NGC 5548.  
Open circles represent the data, from which the HOST$+$NL component is not subtracted, 
and stellate symbol represents the HOST$+$NL component.  
Thick line represents the best-fit linear regression. 
Bottom panel shows the residual plot of the data from the best-fit
regression line. 
The residual here is expressed as a ratio relative to the 
mean observed flux for which the HOST$+$NL component has been subtracted.  
\label{NGC4051ffBV2}}
\end{figure}

\begin{figure}
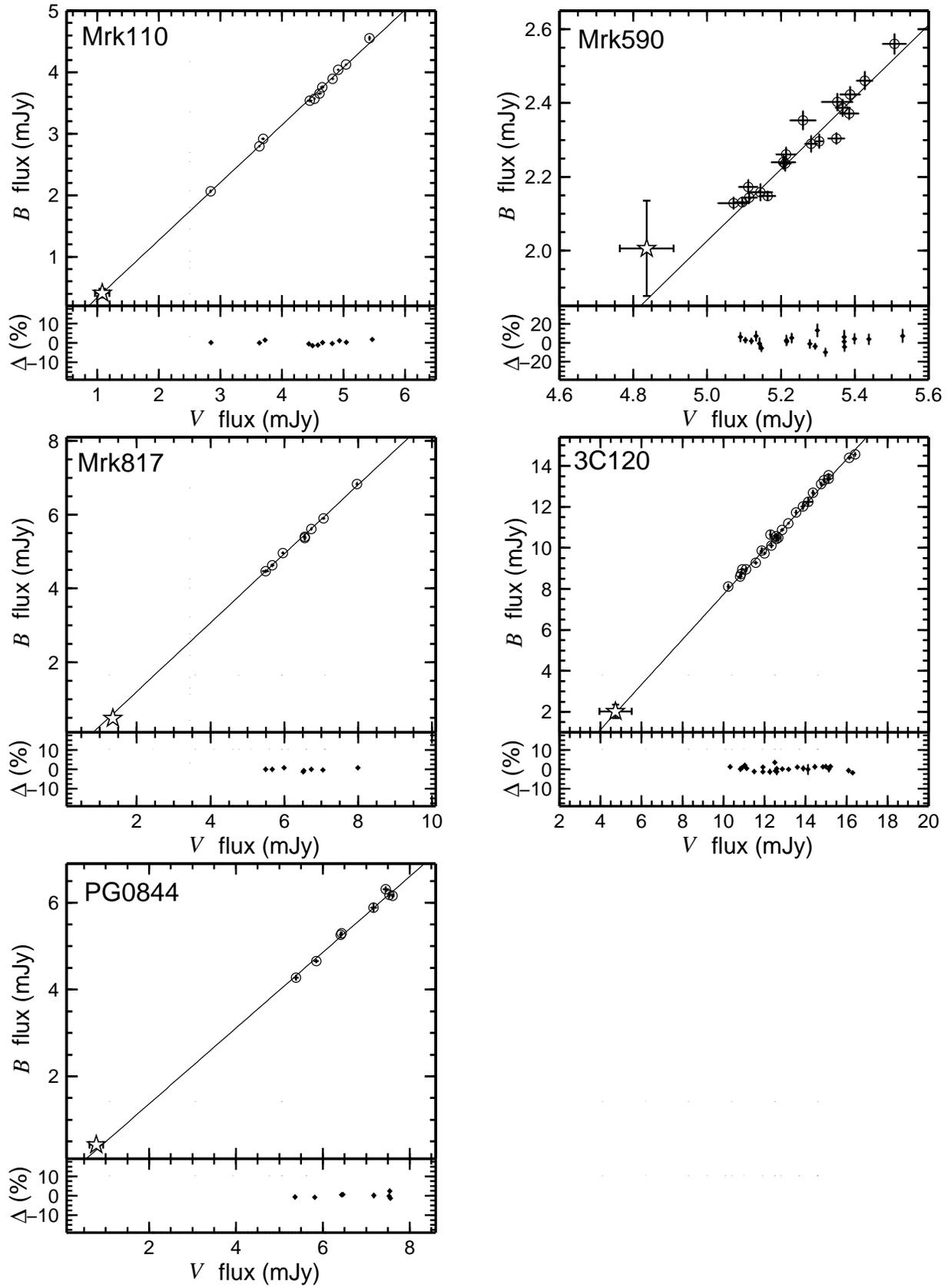


\plottwo{f21a.eps}{f21b.eps}

\plottwo{f21c.eps}{f21d.eps}

\plottwo{f21e.eps}{f21f.eps}

\caption{The same as in Figure 
\ref{NGC4051ffBV2}, but for Mrk 110, Mrk 590, Mrk 817, 3C120 and PG0844.
\label{NGC5548ffBV2}}
\end{figure}

\begin{figure}
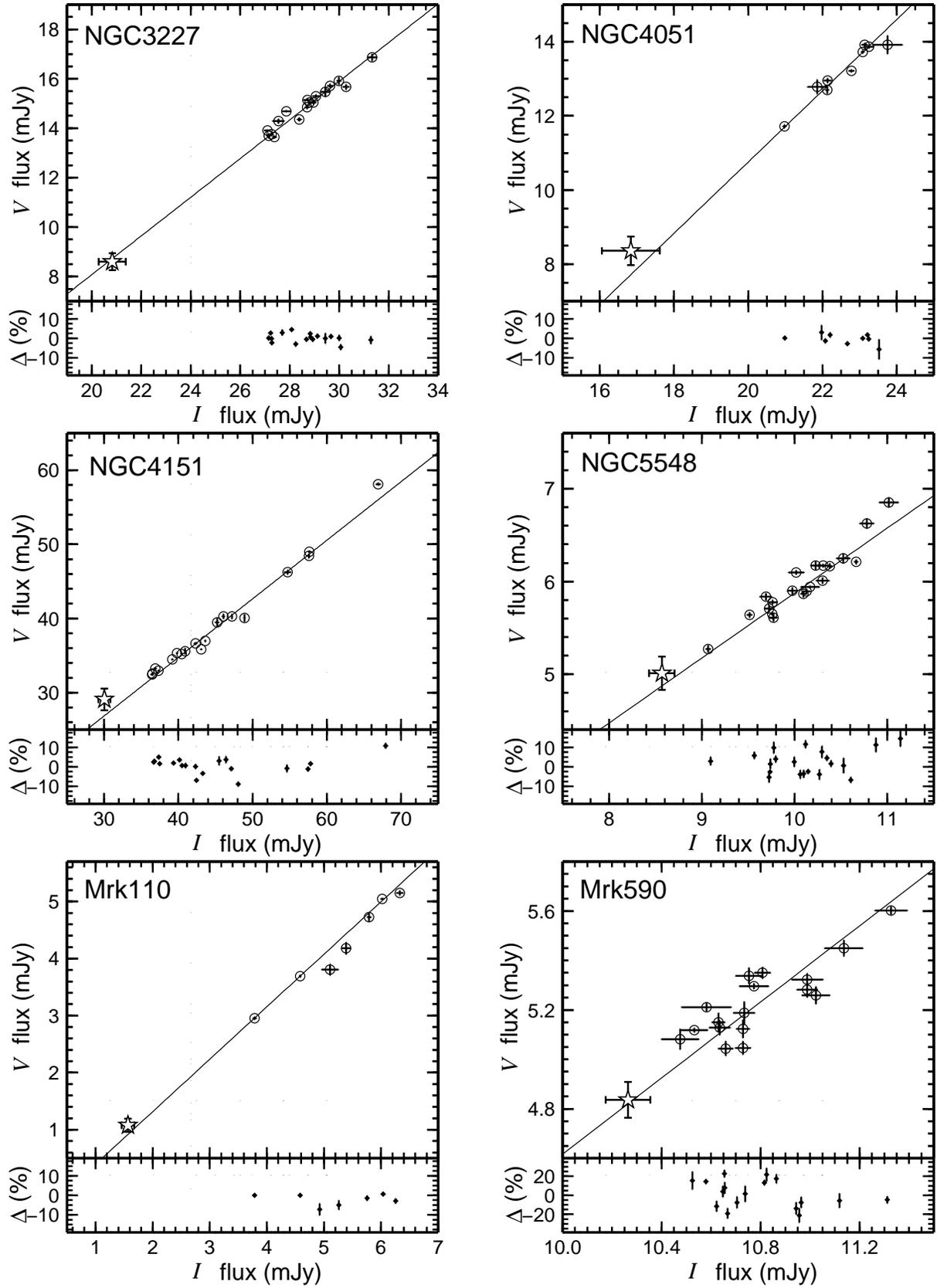


\plottwo{f22a.eps}{f22b.eps}

\plottwo{f22c.eps}{f22d.eps}

\plottwo{f22e.eps}{f22f.eps}

\caption{The $V$-band  flux to $I$-band flux diagrams for target AGNs of 
NGC 3227, NGC 4051, NGC 4151, NGC 5548, Mrk 110 and Mrk 590. 
The same as in Figure \ref{NGC4051ffBV2}.
\label{N3227ffIV2}}
\end{figure}

In this study, we plot only $(B,V)$ and $(V,I)$ pair data in the flux to flux diagram, 
because our monitoring data are largest in the $V$ band, and 
the number of other paired data obtained on the same nights is relatively small.   
$U$-band data were not used because it was difficult to estimate the significant 
contribution of the small blue bump, blended \ion{Fe}{2} lines and Balmer continuum in this band. 
Figures \ref{NGC4051ffBV2} and \ref{NGC5548ffBV2} show the $B$-band flux 
to $V$-band flux diagrams for all 11 targets.   
Figures \ref{N3227ffIV2} and \ref{mrk817ffIV2} show the $V$-band flux to 
$I$-band flux diagrams for 8 targets, because the other 3 targets, 
NGC 3516, NGC 4593, and PG0844+349, were not monitored in the $I$ band. 
Straight-line fitting to the pair data in the flux to flux diagram was 
done with the fitting code made by Tomita (2005) following the most generalized 
multivariate least square process as given by Jefferys (1980, 1981).   
The parameter values of the best-fit linear regression lines, 
the number of the data points, and the reduced $\chi^2$ are listed in Table \ref{linefittable}. 
The residuals of the data from the best-fit linear regression line are plotted 
in the lower panels of the flux to flux diagrams in Figures 
\ref{NGC4051ffBV2} to \ref{mrk817ffIV2}, 
and the ratio of $R_V=\sigma _V^{\rm res} / \Delta V$ is listed in Table \ref{linefittable}, 
where $\sigma _V^{\rm res}$ is the standard deviation of the residuals and 
$\Delta V = f_V^{\rm max}-f_V^{\rm min}$ is the amplitude of flux variation.

As shown in Figures \ref{NGC4051ffBV2} and \ref{NGC5548ffBV2}, 
the $(B,V)$ pair data are certainly distributed in a highly linear manner for all 11 targets, 
and no curvature can be seen in the plots. 
As listed in Table \ref{linefittable}, the residuals of the straight-line fitting of the 
$(B,V)$ plots are so small that the reduced $\chi^2$ is near unity, 
except for two targets of NGC 4151 and Mrk 110.   
Indeed, according to the $\chi^2$-test for the straight-line fitting of the $(B,V)$ plots, 
the linear relationship is not rejected
for seven of the 11 targets at a 1\% level of significance. 
On the other hand, for the rest of the targets, NGC 4051, NGC 4151, NGC 5548, 
and Mrk 110, the $\chi^2$-test rejects the linear relationship at the 1\% level of significance, 
but the ratio of fitting residual to flux variation $R_V$ is comparable to or even less than 
that for the other seven targets.
The $\chi^2$-test and the $R_V$-ratio suggest that
higher photometric accuracy and/or larger number of data could reveal small 
deviation from the linear correlation for these objects. 
We here note that small contamination by broad emission-lines, which show 
time-delayed response to optical continuum emission, is liable to cause a small scatter around 
the linear correlation for the $(B,V)$ plots.

As shown in Figures \ref{N3227ffIV2} and \ref{mrk817ffIV2}, the $(V,I)$ pair data are 
also certainly distributed in a highly linear manner for all eight targets, and no curvature can 
be seen in the plots.  
However, the reduced $\chi^2$-value of the straight-line fitting 
and the $R_V$-ratio of fitting residual to flux variation for the $(V,I)$ plots tend to be 
larger than those for the $(B,V)$ plots. 
Indeed, according to the $\chi^2$-test, the linear relationship for the $(V,I)$ plots is 
rejected  for five of eight targets at the 1\% level of significance. 
The origin of the relatively larger scatter in the $(V,I)$ 
plots will be discussed in Section \ref{effectDT}.

\subsection{Location of Host-Galaxy plus Narrow-Line Component in the Flux to Flux Diagram}
\label{LocationHN}

The location of the host-galaxy plus narrow-line component (HOST+NL component) in the 
flux to flux diagram is presented in Figures \ref{NGC4051ffBV2} to \ref{mrk817ffIV2} 
for the target AGNs.  We define the error of the HOST+NL flux as 
$\sigma = \sqrt{\sigma_{\rm HOST}^2+\sigma_{\rm NL}^2}$, 
where $\sigma_{\rm HOST}$ is the error of the host-galaxy flux 
and $\sigma_{\rm NL}$ is the error of the narrow-line flux, respectively.  
We assume $10$\% for $\sigma _{\rm NL}$,
because the error of the narrow-line flux is not presented
in \citet{1992ApJS...79...49W} but the standard deviation of
several measurements of the narrow-line flux for the individual objects 
in his sample ranges from several \% to a few of $10$ \%.
We note that $\sigma _{\rm NL}$ usually has a minor contribution
to the total error $\sigma$.  
It is evident that the HOST$+$NL component is located on the fainter extension of 
the best-fit linear regression line within 2$\sigma$ error in almost all cases, 
and even in only one exception of Mrk 817 $(V,I)$, the deviation of the HOST+NL 
component from the best-fit linear regression line is still about 3$\sigma$ level. 
We note that the angular size of bulge component of Mrk 817 is so small 
($R_e=0''.5$) that decomposing the bulge and nucleus is relatively difficult.  
Therefore, the linear correlation between the fluxes
in different bands and the location of the HOST+NL
component on the flux to flux diagram strongly indicate that the AGN optical continuum 
does not systematically change the spectral shape while changing the brightness.  
In other words, it does not become bluer even in the brighter phase.

By subtracting the HOST$+$NL component, we arrive at a conclusion that during 
seven years of monitoring observations, the AGN optical continuum flux varied by $0.3$--$3.0$ mag. 
In fact, the optical spectral shape of the AGN continuum emission of NGC 4151 and NGC 5548, 
which showed the largest-amplitude variations, remained nearly constant during flux 
variations of more than 10 times.

\section{Effects of Dust Torus Emission}
\label{effectDT}

While a straight line provides a good fit to almost all $(B,V)$ and $(V,I)$ plots, 
we see noticeable scatter around the line for many of the $(V,I)$ plots.  
This was also pointed out by Winkler (1997), who suspected the existence of some additional 
variable component that could contaminate the $I$-band flux and cause scatter in 
the $(V,I)$ plots.

Thermal emission from dust torus is thought to be the most promising candidate for 
an additional variable component that could contaminate the optical flux at longer wavelengths.  
This emission in the $I$ band could then disturb the linear relationship in the $(V,I)$ plots 
because it shows a time-delayed response to emission in the same band from the 
accretion disk (Minezaki et al. 2004; Suganuma et al. 2006).  
We here estimate the flux contribution of dust torus in the $I$ band, 
using simultaneous $VIJ$ light curves obtained by the MAGNUM telescope. 

First, we assume that the flux in each band is decomposed into accretion disk, HOST$+$NL, 
and dust torus, while the emission from dust torus in the $V$ band is ignored.  
That is because the sublimation temperature of dust is about $1800$ K and does not exceed 
$2000$ K \citep{1977ARA&A..15..267S,1999IAUS..191..279L}, and blackbody radiation 
at $2000$ K in the $V$ band is less than 5$\%$ of that in the $I$ band.  
Then, the fluxes in the $VIJ$ bands are expressed as
\begin{eqnarray}
 f_V(t) &=& f_V^{AD}(t)+f_V^{HN} \;\;,  \\
 f_I(t) &=& f_I^{AD}(t)+f_I^{HN}+f_I^{DT}(t) \;\;,  \\
 f_J(t) &=& f_J^{AD}(t)+f_J^{HN}+f_J^{DT}(t) \;\;, 
\label{emissionassumption}
\end{eqnarray}
\\
where the superscript AD stands for accretion disk, HN for HOST$+$NL, and DT for dust torus. 
Second, we assume that the $V$ to $J$ flux ratio in the power-law spectral emission from 
the accretion disk is kept nearly constant irrespective of its flux variation, namely
\begin{equation}
f_V^{AD}(t)/f_J^{AD}(t) = \left( \frac{\nu_V}{\nu_J} \right)^{\alpha_{VJ}}  \;\;,
\label{VKADratio}
\end{equation}
\\
where $\nu_V$ and $\nu_J$ are the effective frequencies in the MAGNUM $V$ and $J$ filters, 
respectively, and $\alpha_{VJ}$ is the power-law spectral index of constant value.  
Finally, although the constant color of $I-J$ and the synchronized variation in the $I$ and 
$J$ bands for dust torus are still under discussion, we assume that the $I$ to $J$ flux ratio 
in the thermal emission from dust torus is also kept constant, namely
\begin{equation}
f_I^{DT}(t)/f_J^{DT}(t) = \left( \frac{\nu_I}{\nu_J} \right)^{\beta}\frac{B_{\nu_I}(T_{d})}{B_{\nu_J}(T_{d})}  \;\;,
\label{IKdustration}
\end{equation}
\\
where the power index of $\beta$ has a constant value of 1.6 from the absorption efficiency 
$Q_{abs}$ for graphite particles of radius $0.05\mu m$ in the relevant wavelength region 
(Draine \& Lee 1984), and $B_{\nu}(T_d)$ is the specific intensity of a black body of 
constant dust temperature $T_d$ at frequency $\nu$.  
We note that a single temperature is assigned to dust torus, 
which is realistically of a multi-temperature component.  

With the above assumptions, the $I$-band flux from dust torus is written as 
\begin{eqnarray}
f_I^{DT}(t)=&(&f_J(t)-(f_V(t)-f_V^{HN})(\nu_J/\nu_V)^{\alpha_{VJ}}-f_J^{HN}) \nonumber  \\
&\times&(\nu_I/\nu_J)^{\beta}(B_{\nu_I}(T_{d})/B_{\nu_J}(T_{d})) \;\;,
\label{Idustcomponenteq}
\end{eqnarray} 
where $f_V(t)$ and $f_J(t)$ are the observed light curves. 
$f_V^{HN}$  was estimated in Section \ref{LocationHN}, 
and $f_J^{HN}$ was estimated as described in Section \ref{nonvarcompestimate} 
using MAGNUM $J$-band images (the narrow-line contribution in the $J$ band was ignored).  
Consequently, two parameters, $T_d$ and $\alpha_{VJ}$, remain to be determined.  
In other words, given these parameters, we can subtract the contribution of dust torus 
$f_I^{DT}(t)$ from the observed $I$-band flux and fit a straight-line 
to the $(V,I)$ plots corrected for dust torus emission. 

We searched for the best set of these parameters that could minimize the $\chi^2$-value 
for the straight-line fitting to such dust-free $(V,I)$ plots by repeating the $\chi^2$  
calculation from $T_d=1000$ K to 3000 K at intervals of 50 K and from $\alpha_{VJ}=-3$ 
to $+3$ at intervals of $0.1$.  
For example, Figure \ref{n4151aTchi} shows the contour map of reduced 
$\chi ^2$ for the parameter search for NGC 4151.  
Figures \ref{Idust4051} and \ref{Idust110} present the $V$-band flux 
to $I$-band flux diagram and the best-fit linear regression line with and without 
the dust correction for seven targets. 
Mrk 590 was removed from the discussion 
because its AGN component was so much fainter than the host galaxy in the $J$ band 
that accurate image decomposition was difficult.  
The residuals of the data from the best-fit linear regression line are also plotted 
in the middle and lower panels of Figures \ref{Idust4051} and \ref{Idust110}. 
The best-fit parameters of $T_d$ and $\alpha_{VJ}$ with the $J$-band host-galaxy 
flux $f_J^{HN}$ are tabulated in Table \ref{Idustcontritable}, together with 
the best-fit $\alpha_{VI}$ and the reduced $\chi ^2$ of the straight-line 
fitting to the $(V,I)$ plots after subtracting the dust component in the $I$-band.  
For reference, the best-fit $\alpha_{VI}$ and the reduced $\chi ^2$ without 
the dust correction are also tabulated in Table \ref{Idustcontritable}.  

As a result of the correction for contamination by dust torus emission in the $I$ band, 
we found that the reduced $\chi^2$-values for all target AGNs dramatically 
decreased to $\chi^2 \sim 1$ for six of the seven targets except for NGC 4151. 
Indeed, the linear relationship for the $(V,I)$ plots
after the dust correction turned out to be 
not rejected
at the 1\% level of significance, 
and even for NGC 4151, the reduced $\chi^2$-value decreased remarkably 
with the dust correction.  
We also found that the location of the HOST+NL component in the 
$(V,I)$ plot became closer to the fainter extension of the best-fit regression line 
after the dust correction in almost all cases.  
In addition, the best-fit value of $T_d$ ($1700 \sim 2000$ K) 
became consistent with the upper limit of the sublimation temperature of dust grains.  
On the other hand, the best value of $\alpha_{VJ}$ ($-0.7\sim 0.0$) was generally smaller 
than the value reported by Tomita (2005).  
Most values of $\alpha_{VJ}$ showed good agreement with 
the value of $\alpha_{VI}$, but not for NGC 5548 and Mrk 817. 
However, $\alpha_{VJ}$ would not be constrained well 
because the reduced $\chi^2$-value was insensitive to any change of $\alpha_{VJ}$.
The reduced $\chi^2$-value for Mrk817 was so small that 
the linear relation seemed to provide a satisfactory fit without the dust correction
\footnote{
We note a small number of the $(V,I)$
points available for Mrk 817.
}, 
however, it also much decreased by the dust correction, 
and the best-fit $T_d$ showed a reasonable value.  
This suggests that 
the $I$-band flux is contaminated 
by a dust component for Mrk817 as well.

From these exercises, we conclude that the contribution of dust torus to the optical flux at 
longer wavelengths cannot be negligible in discussing the optical color variability, and 
most of the scatter around the best-fit regression line in the $(V,I)$ plot can be 
explained by contamination of dust torus emission to the observed $I$-band flux.  
Along with the results of Section \ref{testcolorvarchap}, this result confirms that 
the optical spectral shape ($0.44 \sim 0.79$ $\mu$m) of AGNs remains nearly constant 
during their flux variation.       

\section{Discussion}
\label{discusschap}

\subsection{Spectral Variability of UV-Optical Continuum Emission in Previous Studies}

The results of preceding sections showed that AGN optical continuum does not 
systematically change in spectral shape during flux variation.  
In this section, we review the previous studies of AGN spectral 
variability in the UV-optical region and discuss them in comparison with our results.  

\subsubsection{Optical Color Variability by \citet{2000ApJ...540..652W} and \citet{1999MNRAS.306..637G}}

We examine two leading studies that claimed spectral hardening of
intrinsic emission from AGNs, opposite to our conclusion.  

\citet{2000ApJ...540..652W} monitored 23 AGNs in the blue, yellow, and 
red bands for three months.  
They constructed a spectral variation model of AGN plus host galaxy 
with the AGN component contributing two thirds of the total flux in the yellow band 
as an extreme case for a large contribution of host galaxy.
Then, allowing 0.5 mag variation for the AGN component, 
they estimated the amplitude of flux variation in the blue, yellow and red bands,  
and found, as in their Figure 8, that the blue to red amplitude ratio exceeded that expected 
from the spectral variation model for two of six targets. 
Accordingly, they concluded that the intrinsic emission from AGNs 
became bluer when they were brighter.

\citet{2000ApJ...540..652W} estimated the host-galaxy contribution 
for more than two thirds of their objects.
Since
four of them are included in our sample,
it is instructive to compare their estimation of host-galaxy flux with our estimation 
with improved accuracy as described in Section 3.3.  
Their host-galaxy flux in the red band is about
13 mJy (NGC 3227), 8.3 mJy (NGC 4151), 3.0 mJy (NGC 5548), and 1.4 mJy (Mrk 817), 
with an absolute calibration error of $20-30$\%.  
Their aperture diameter of $11''$ is larger than ours of $8''.3$, 
but their estimation of host-galaxy flux is much smaller than ours in the $I$ band
(the effective wavelength of their red band is close to that of the $I$ band).
Moreover, their assumption of the host-galaxy component contributing 
one third of the total flux in the yellow band is not at all
an extreme case. 
Table \ref{hostcont} shows the contribution of the HOST$+$NL component
in the average flux of our light curve data.  We see that the contribution 
in the $V$ band, for which the effective wavelength is slightly shorter than 
that of the yellow band, exceeds one third for eight out of 11 targets,
and even two thirds for 5 targets.
Therefore, the targets, which show a large blue to red variation amplitude ratio 
in \citet{2000ApJ...540..652W}, 
do not necessarily indicate the spectral hardening of
intrinsic emission from AGNs.

Giveon et al. (1999) monitored 42 PG quasars in the $BR$ bands for seven years, 
which were generally more luminous than our targets.  
Their photometry was based on PSF fitting, and the host-galaxy contribution 
was not subtracted.
They found that at least half of their targets 
showed bluer $B-R$ color when they became brighter.
It is therefore worthwhile to examine the color variability
of their targets by means of linear correlation analysis described in Section 5.

First, for PG 0844+349 that is the only target in common with ours,
a tight linear relationship for the $(B,V)$ plot is presented and  
the HOST$+$NL component is found to be located on the fainter extension 
of the best-fit linear regression line.  
Second, for six targets that were more luminous 
than PG 0844+349 by about 1.5 mag in the $B$ band and that showed significant 
flux variation of $\Delta B>0.6$ mag (see their Table 3),  
their $BR$ monitoring data are plotted in a flux to flux diagram as shown in 
Figure \ref{PG0953}, together with the best-fit regression line.  
We see that the linear relationship is a good approximation for the six targets, 
in disagreement with the conclusion by \citet{1999MNRAS.306..637G} 
that they all became bluer when they became brighter.  
The scatter around the best-fit regression line in Figure \ref{PG0953} is somewhat larger 
than that for our targets shown in Figures \ref{NGC4051ffBV2} to \ref{mrk817ffIV2}.  
Such scatter would be caused by the broad emission-line 
components of small blue bump and Balmer lines that contaminate the flux 
in the $B$ and $R$ bands, and also by the uncertainty associated with 
PSF-fitting photometry.  
The good linear relationship for the $(B,R)$ plots demonstrates that the 
optical spectral shape of the variable component remains nearly constant 
even for the luminous QSOs. Although it is not examined whether or not
the HOST$+$NL component of them locates on 
the fainter extension of the linear regression line, by combining our results 
for Seyfert galaxies and PG 0844$+$349,
it is strongly indicated that optical continuum of AGNs from Seyfert galaxies 
to QSOs does not systematically change in the spectral shape during flux variation.  

\subsubsection{UV Color Variability}

Seyfert 1 galaxies and QSOs produce strong continuum emission in the UV region 
as well as in the optical region.
Similarly to the optical color variability, two opposite claims on the UV 
color variability 
have prevailed without reaching a conclusion as to whether or not the 
UV continuum of AGNs changes in spectral shape during flux variation. 

Paltani \& Walter (1996) observed the UV spectra of 15 nearby AGNs 
at different epochs using the IUE satellite.  
By principal component analysis, they explained UV variation in almost all their targets 
as the sum of a variable component of constant power-law spectral shape, 
with evidence of reddening by dust extinction, and a non-variable component 
of small blue bump consisting of  
steep Balmer continuum and \ion{Fe}{2} pseudo-continuum. 
Other studies arriving at similar results include Rodriguez-Pascual et al. (1997) 
and Santos-Lleo et al. (1995), who observed Fairall 9 and NGC 4593, 
respectively, using the IUE satellite. 
From discussions on the linearity of UV monitoring data in the 
$f(1803 \AA)$ to $f(1447 \AA)$ diagram, they concluded the constant 
spectral shape in the UV region during their flux variation.  
Santos-Lleo et al. (1995) also estimated the non-variable 
component for NGC 4593 and found that it is located on the fainter extension 
of the best-fit regression line within the uncertainties. 

On the other hand, for NGC 5548 (Wamsteker et al. 1990; Korista et al. 1995), NGC 4151 
(Crenshaw et al. 1996), and NGC 3783 (Reichert et al. 1994), the claim of spectral hardening 
in the UV region was argued from the changeover of the power-law spectral index through 
flux variation, or from the larger amplitude of variability at shorter wavelengths.  
Although most of these studies did not take into account the effect of non-variable component, 
Wamsteker et al. (1990) considered the host-galaxy component. 
Romano \& Peterson (1998) plotted quasi-simultaneous UV and optical monitoring data for 
NGC 5548 in the $F(1350 \AA)$ to $F(5100 \AA)$ diagram.  
They found a positive correlation with a curvature to be statistically consistent 
with a linear correlation in the $F(1350 \AA)$ - $F(5100 \AA)$ plots, hence 
they argued for a change in spectral shape. 

Vanden Berk et al. (2004) studied the relationship between variability amplitude and rest-frame 
wavelength of the UV-optical region from SDSS two-epoch multicolor observations of 
about 25,000 quasars and found the larger amplitude at shorter wavelengths 
in the UV region of $\lambda < 4000 \AA$, although this trend seemed to be very small, 
and the amplitude was almost constant in the optical region of 
$4000 \AA < \lambda < 6000 \AA$, as presented in their Figure 13.  
In their study, the non-variable component was not subtracted, but they noted that 
if its contribution had been significant, the variability amplitude should have 
decreased drastically at $\lambda > 4000 \AA$, which would simply 
strengthen the correlation they found.  
Therefore, they concluded that quasars were systematically bluer when brighter.  
On the other hand, their results in the optical region of $4000 \AA < \lambda < 6000 \AA$, 
which overlaps with our $B$-band and $V$-band wavelength regions, show little or no change in 
amplitude, and in fact, they are consistent with our conclusion of nearly constant 
spectral shape in the optical region.

Although a problem is still controversial as to whether or not the UV spectral shape 
really changes, it is possible to conceive a theoretical model that allows 
spectral hardening in the UV region and constant spectral shape in the optical region.  
Recently, \citet{2006ApJ...642...87P} presented such a model, assuming the standard 
accretion disk \citep{1973A&A....24..337S} and variation of mass accretion 
rate onto the disk.  
The variation of accretion rate changes the maximum disk temperature
without changing the radial temperature profile in the disk.  
Thus, the UV spectrum would become harder when it brightens,
because it originates from the innermost disk
where the maximum temperature increases with the mass accretion rate.  
On the other hand, the optical spectrum would remain nearly constant,
because it originates from relatively outer region of the accretion disk
where the temperature structure remains unchanged.
Figure 1 of \citet{2006ApJ...642...87P} shows that the model predicts 
large spectral variation in the UV region while little spectral variation in the optical region.  

In addition, the color variability in the UV region might reflect different physical 
properties of individual AGNs.   
UV spectral hardening was reported for some AGNs
(NGC 5548, NGC 4151, NGC 3783), while nearly constant spectral shape was reported 
for other AGNs (Fairall 9, NGC 4593).  
According to the model for flux variation above, the spectral shape of the UV continuum
changes with the maximum disk temperature, which is dependent
on the mass accretion rate and the central black hole mass.
In order to settle this problem, highly accurate simultaneous UV and optical monitoring 
observations are necessary for many AGNs, together with high-resolution images 
for estimation of their non-variable component.      

\subsection{Optical Color of Accretion Disk}

The slope of the best-fit regression line in the flux to flux diagram 
corresponds to the color of variable AGN component.  Our conclusion 
in the preceding sections is that the optical spectral shape of AGNs 
remains nearly constant during flux variation, hence the optical color derived from the 
flux to flux diagram can be regarded as the color of the AGN optical continuum.  
Thus, the color of AGN optical continuum between the $X$ and $Y$ bands is derived as
\begin{equation}
X-Y = -2.5 (\log_{10}(a_{XY})-\log_{10}(f_{X0}/f_{Y0})) \;\;,
\label{colorXY}
\end{equation} 
where $a_{XY}$ is the slope of the best-fit regression line in the $X$-band flux to 
$Y$-band flux diagram, 
and $f_{X0}$ and $f_{Y0}$ are the fluxes at zero magnitude in the respective bands.  
Assuming the power-law spectrum of $f_{\nu} \propto \nu^\alpha$ for the AGN 
optical continuum, we obtain the power-law spectral index $\alpha$ between 
the $X$ and $Y$ bands as
\begin{equation}
\alpha_{XY} = \frac{\log(a_{XY})}{\log(\nu_X/\nu_Y)} \;\;,
\label{alphaXY}
\end{equation}  
where $\nu_X$ and $\nu_Y$ are the effective frequencies of the respective bands.  

Table \ref{variablecolor} shows the color and spectral index obtained from 
the best-fit regression line of the flux to flux plots.  
They are compared with those obtained by \citet{1997MNRAS.292..273W} 
and \citet{2001AJ....122..549V}, as shown in Figure \ref{varcolalpha}.  
Furthermore, since the AGN optical continuum emission is considered to originate from the 
accretion disk \citep{1978Natur.272..706S,1982ApJ...254...22M}, 
the spectral index expected from the standard accretion disk 
\citep{1973A&A....24..337S} at long wavelengths with an infinitely 
extended disk is also shown in this figure for comparison.  
We found that the colors are comparable to 
that of standard accretion disk or slightly redder, and are bluer 
than the QSO composite color by \citet{2001AJ....122..549V}. 

Winkler (1997) plotted the optical colors of the variable component of 92 Seyfert 1 galaxies 
in the two-color diagram, and found that most targets were distributed around the bluest target, 
while redder targets were distributed in the direction of the reddening vector.  
He interpreted this result to mean that the bluest color was the intrinsic color of Seyfert galaxies, 
and the redder colors were caused by internal extinction from their circumnuclear regions.  
The $(B-V)$ to $(V-I)$ diagram is shown in Figure \ref{varcolcol}, where the data of 
Winkler (1997) are plotted together with our data.  
As shown in Figure \ref{varcolcol}, our data present a similar distribution with those of Winkler (1997).  
Following his method, we estimated the internal extinction of redder targets and compared 
the extinction values with those estimated in previous studies. 

Paltani et al. (1996) estimated the internal extinction by fitting the power-law 
spectral emission to the UV continuum for nearby Seyferts, and some of their targets 
were in common with ours.  
Crenshaw et al. (2001) estimated the internal extinction of NGC 3227 by 
comparing its UV spectrum with that of NGC 4151. 
We then compared our estimates of extinction 
with theirs and found them consistent with each other, 
as shown in Table \ref{internalextinctioncomparison}. 
The consistency between the reddening of optical colors obtained from the flux to flux 
plots and the extinction estimated from other methods supports the idea of Winkler (1997) 
that the optical color of the variable component, 
after corrected for internal extinction [$E(B-V)\simeq0$--$0.2$mag], is interpreted as the 
intrinsic color, which is similar to the color of standard accretion disk.

\section{Summary}
\label{conclusionchap}

We carried out accurate and frequent monitoring observations of 11 nearby AGNs 
(9 Seyfert 1 galaxies and 2 QSOs) in the $B$, $V$, and $I$ bands for seven years 
with the MAGNUM telescope.  
We estimated the flux contribution of the non-variable HOST$+$NL component in the AGN optical 
continuum emission by applying surface brightness profile fitting to HST/ACS high-resolution 
images and MAGNUM images for the host galaxy (HOST), and by analyzing the results of previous 
 spectral observations for narrow emission lines (NLs).  
We found a tight linear flux to flux relationship for the $(B,V)$ and $(V,I)$ plots observed 
on the same days for each target AGN and also found that the HOST$+$NL component 
estimated above could be located on the fainter extension of the best-fit regression line 
in the flux to flux diagram within 2$\sigma$ uncertainties in almost all cases.  
These results strongly support the idea that AGN optical continuum emission does not 
systematically change its spectral shape with changing flux.  
By subtracting the contribution of the HOST$+$NL component from the observed flux, 
we estimated the accurate variability amplitude of AGN optical continuum emission 
and found that some AGNs (NGC 4151, NGC 5548) changed their optical brightness 
by more than 3 mag during the seven-year monitoring period.  
This means that some AGNs show optical flux variation by more than 10 times with their color 
kept nearly constant on a timescale of several years.  
Therefore, spectral hardening, which was argued otherwise in many studies, can be 
interpreted as an apparent trend due to contamination of the non-variable HOST$+$NL component 
in the optical region of $4400 \AA$--$7900 \AA$ considered.  
The nearly constant optical color of the variable component of AGNs is comparable to 
the color of the standard accretion disk or slightly redder.  
The plots of all target AGNs in the $(B-V)$ to $(V-I)$ diagram show that redder targets 
are distributed in the direction of the reddening vector, starting from the bluest color of 
variable component in the diagram, which supports Winkler's idea of the existence of 
an intrinsic AGN color.  
From statistical discussion of straight-line fits, it is found that the scatter around 
the best-fit regression line in the $(V,I)$ plots tends to be larger than that in the $(B,V)$ plots.  
The origin of such large scatter only in the $(V,I)$ plot most likely could be attributed to 
the contamination from dust torus emission in the $I$ band.  
All of these results will make rapid progress in constraining the mechanism of AGN variability 
in the future.        

\acknowledgments

We thank colleagues at the Haleakala Observatories for their help for the facility maintenance. 
This research has been supported partly by the Grant-in-Aid of Scientific Research (10041110, 
 10304014, 12640233, 14047206, 14253001, and 14540223) and COE Research (07CE2002) 
of the Ministry of Education, Science, Culture and Sports of Japan.

\clearpage

\clearpage

\clearpage

\clearpage

\clearpage

\clearpage

\clearpage

\clearpage

\clearpage

\clearpage

\clearpage

\clearpage

\clearpage

\begin{figure}

\plottwo{f23a.eps}{f23b.eps}

\caption{The same as in Figure \ref{N3227ffIV2}, but for Mrk 817 and 3C120. 
.
\label{mrk817ffIV2}}
\end{figure}

\clearpage

\begin{figure}
\plotone{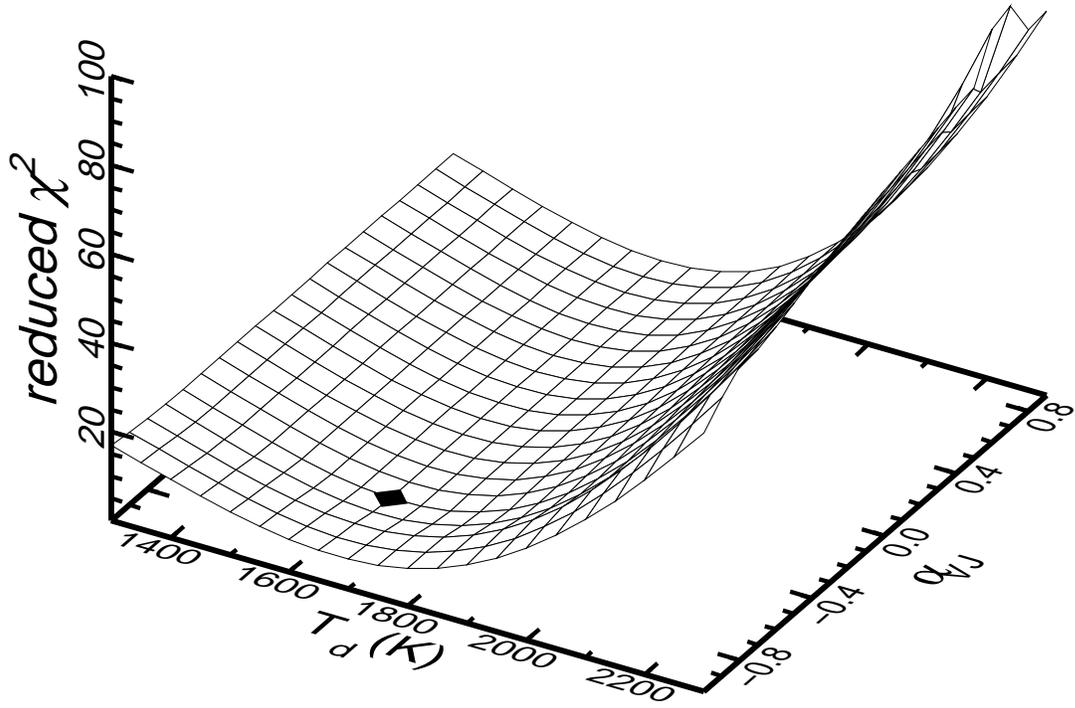}
\caption{The reduced $\chi^2$ map for NGC 4151 as a function of dust temperature $T_d$ and 
power-law spectral index $\alpha_{VJ}$. 
A symbol of filled diamond represents the location at which $\chi^2$ attains a minimum value.
\label{n4151aTchi} }
\end{figure}

\clearpage

\begin{figure}
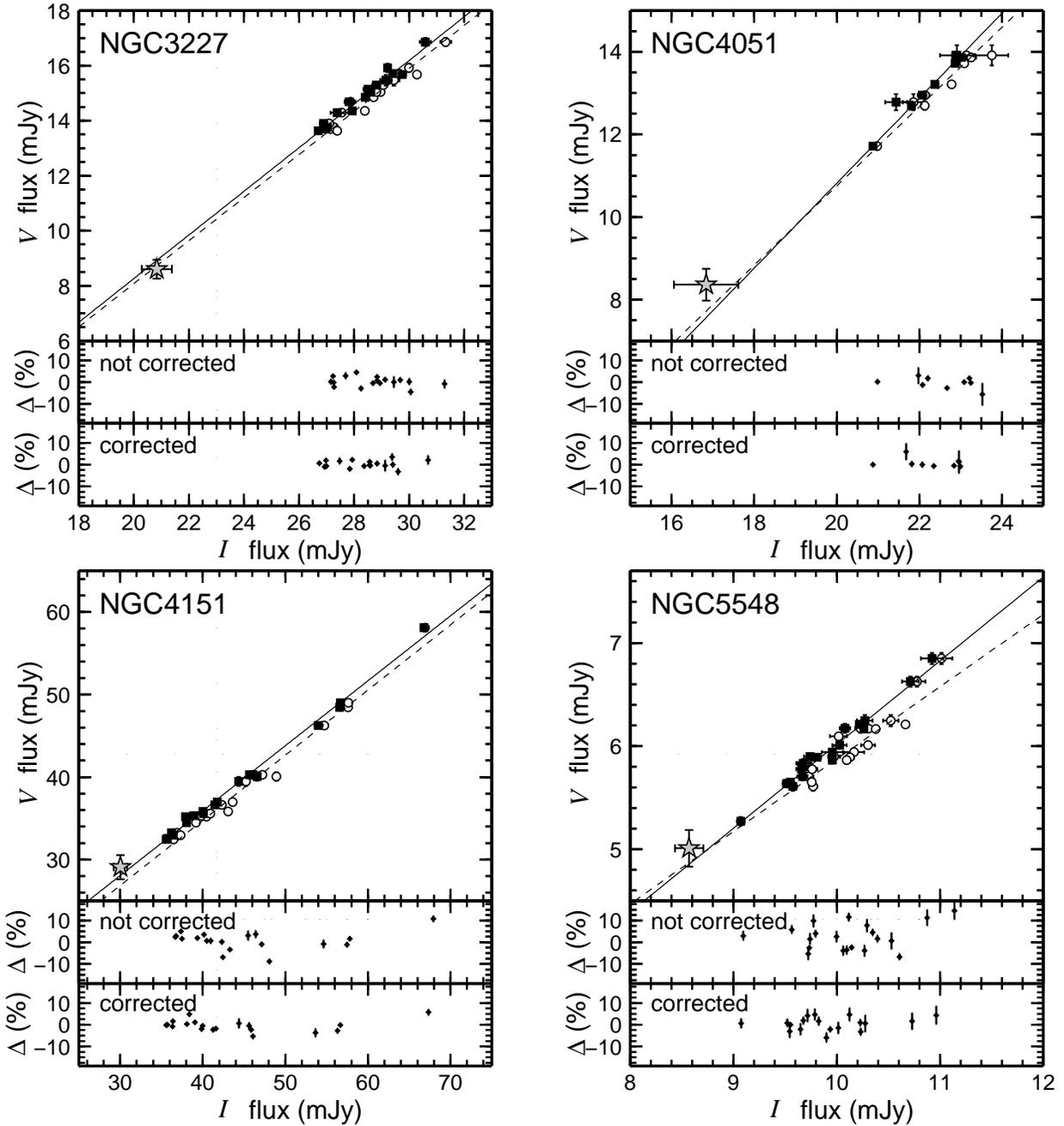


\plottwo{f25a.eps}{f25b.eps}

\plottwo{f25c.eps}{f25d.eps}

\caption{The $V$-band flux to $I$-band flux diagrams 
before and after the correction for dust torus emission for target AGNs of 
NGC 3227, NGC 4051, NGC 4151 and NGC 5548.  
Open circles and dashed line show the data and the best-fit linear regression, 
respectively, before the correction.  
Filled squares and thick line show the data and the best-fit linear regression, 
respectively, after the correction.  
The stellate symbol represents the non-variable HOST+NL component. 
Middle and bottom panels show the residual plot of the data from 
the best-fit linear regression before and after the correction 
for dust torus emission, respectively. 
The residual here is expressed as a ratio relative to the 
mean observed flux for which the HOST$+$NL component has been subtracted.  
\label{Idust4051}
}
\end{figure}

\clearpage

\begin{figure}
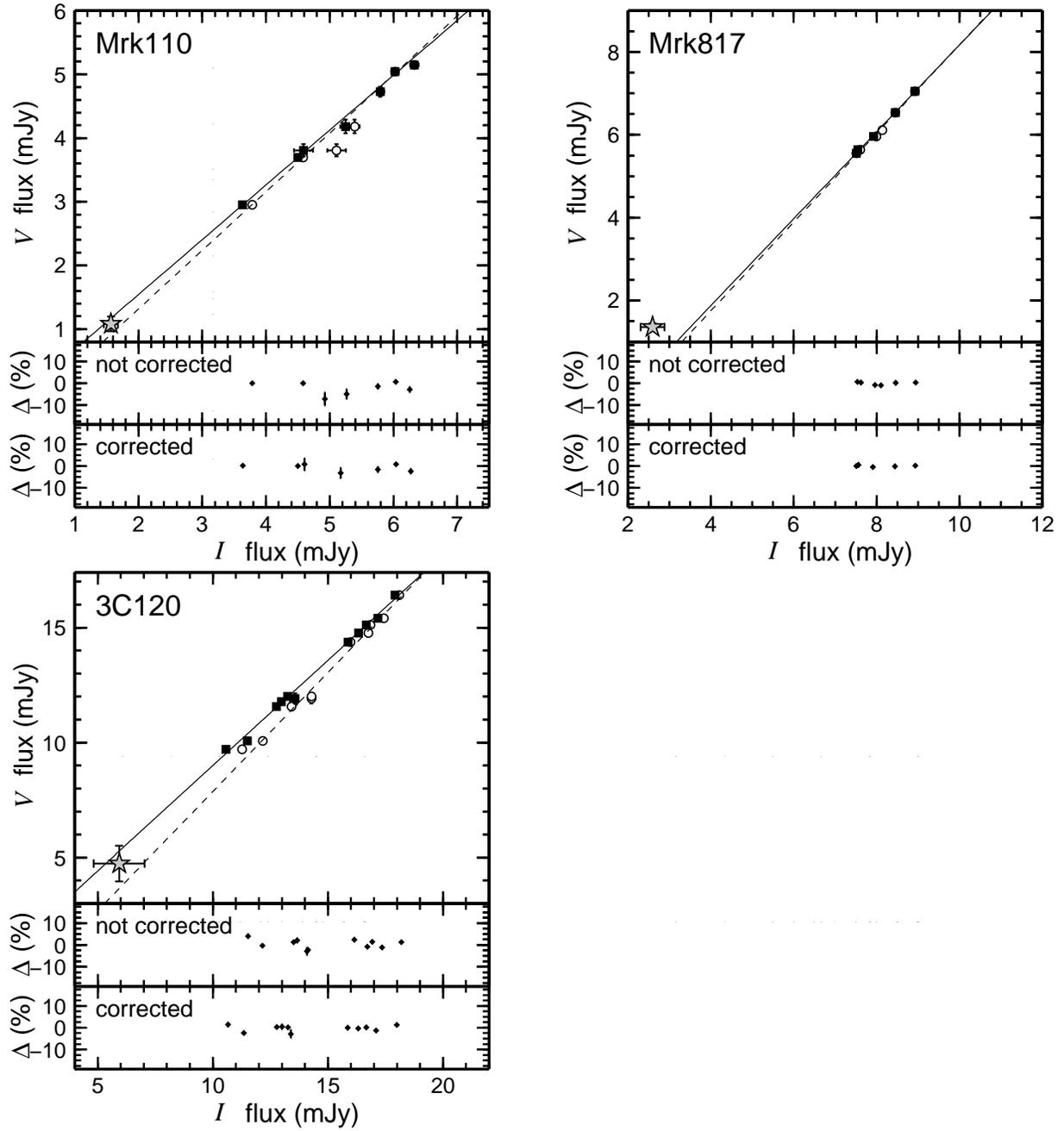


\plottwo{f26a.eps}{f26b.eps}

\plottwo{f26c.eps}{f26d.eps}

\caption{ The same as in Figure \ref{Idust4051}, but for Mrk 110, Mrk 817 and 3C120.
\label{Idust110}
}
\end{figure}

\clearpage

\begin{figure}
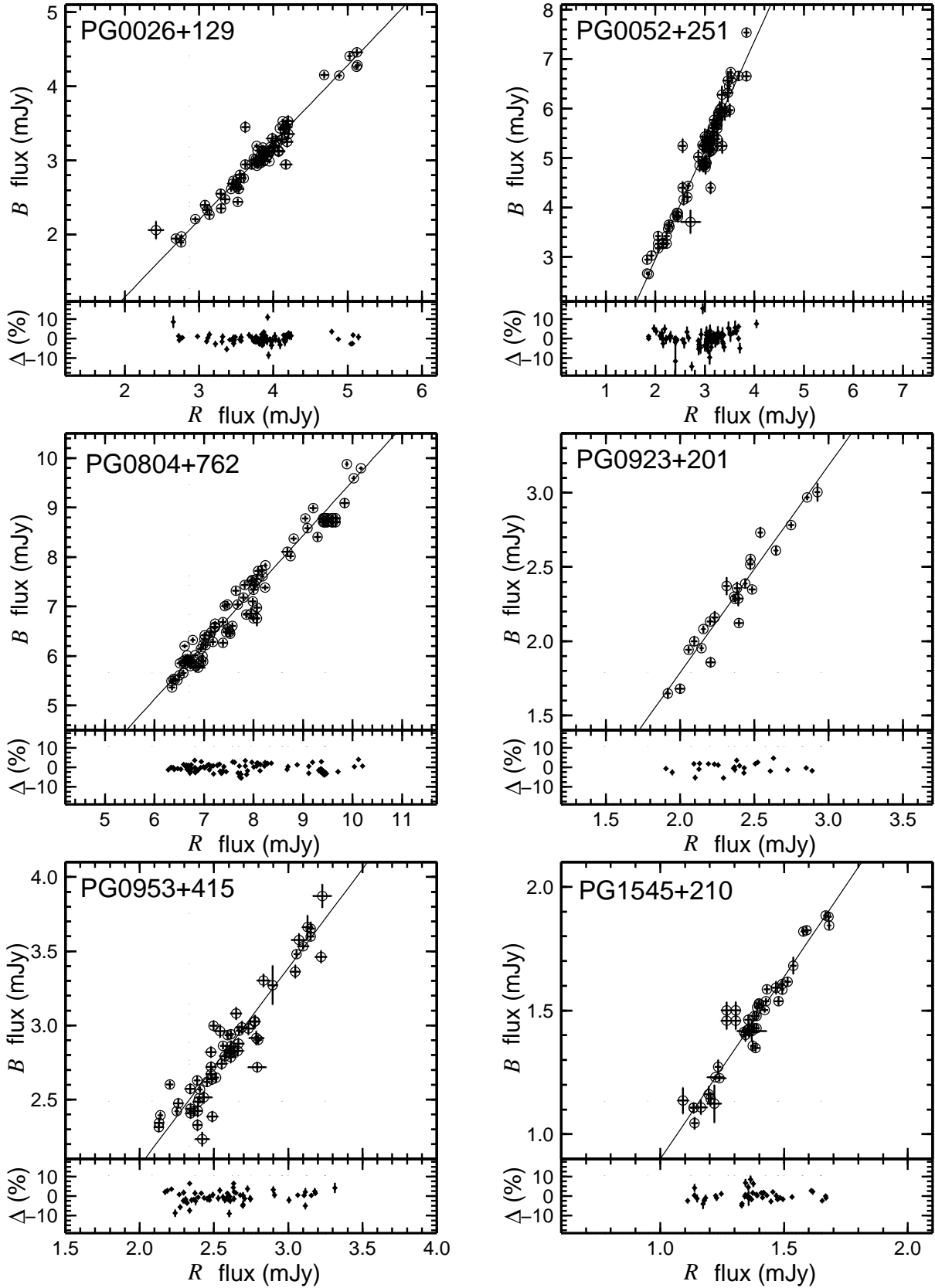


\plottwo{f27a.eps}{f27b.eps}

\plottwo{f27c.eps}{f27d.eps}

\plottwo{f27e.eps}{f27f.eps}

\caption{The $B$-band  flux to $R$-band flux diagrams for 
PG0026+129, PG0052+251, PG0804+762, PG0923+201, PG0953+415,
and PG1545+210, taken from Giveon et al. (1999).  
These are luminous targets in their sample of PG quasars.  
Thick line represents the best-fit linear regression. 
Bottom panel shows the residual plot of the data from the best-fit
regression line. 
The residual here is expressed as a ratio relative to the 
mean observed flux.  
\label{PG0953}}
\end{figure}

\clearpage

\begin{figure}
\epsscale{.80}
\plotone{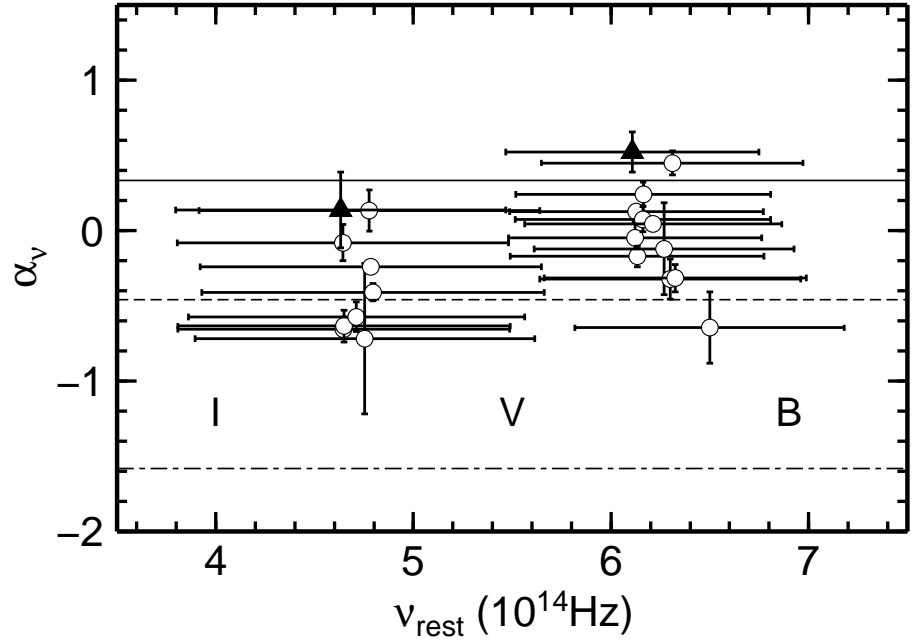}
\caption{The power-law spectral index of AGN continuum plotted 
against rest-frame frequency.  
Open symbols are our results and filled symbols are the mean colors of 
variable component of some Seyfert galaxies by Winkler (1997).  
The solid horizontal line represents the value of $\alpha_\nu=1/3$ expected 
from the standard accretion disk \citep{1973A&A....24..337S}.  
The broken and dash-dotted horizontal lines represent the fitted index of 
QSO composite spectrum in the UV region and in the optical region, respectively
(Vanden Berk et al. 2001).\label{varcolalpha}}
\end{figure}

\begin{figure}
\epsscale{.80}
\plotone{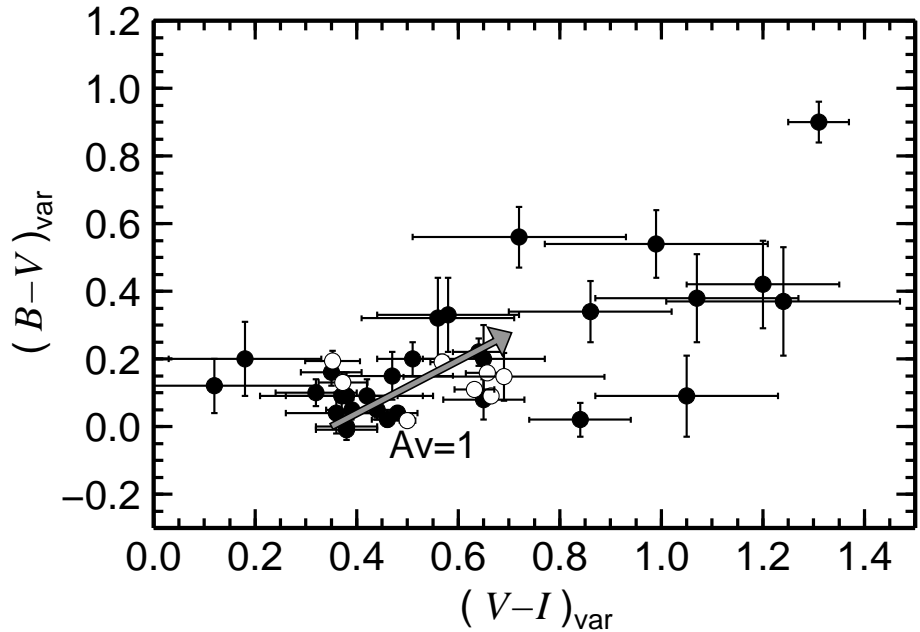}
\caption{ The $(B-V)$ to $(V-I)$ diagram for AGN continuum.  
Open symbols are our results and filled symbols are the results by Winkler (1997).  
The arrow indicates the reddening vector of $A_V=1$ from the bluest color of 
intrinsic AGN color presented in Winkler (1997).
\label{varcolcol}
}

\end{figure}

\clearpage

\begin{deluxetable}{lccccclcl}
\tabletypesize{\footnotesize}
\tablecaption{Characteristics Of Target AGNs \tablenotemark{a}\label{objbasicparm}}
\tablewidth{0pt}
\tablehead{
\colhead{Object} & \colhead{R.A.} & \colhead{Dec} & \colhead{Redshift} & \colhead{Galaxy size} & \colhead{$A_V$} &  \multicolumn{3}{c}{Morphology} \\
\cline{7-9}\\
\colhead{} & \colhead{(J2000.0)} &\colhead{(J2000.0)} &\colhead{} &\colhead{($'$)} &\colhead{(mag)} &\colhead{Host}&&\colhead{AGN}
}
\startdata
NGC3227 & 10h23m30.6s & $+$19d51m54s & 0.003859   & $5.4\times3.6$  & 0.075 & SAB/pec&&Sy1.5\\
NGC3516 & 11h06m47.5s & $+$72d34m07s & 0.008836   & $1.7\times1.3$  & 0.140 & SB0&&Sy1.5\\
NGC4051 & 12h03m09.6s & $+$44d31m53s & 0.002336   & $5.2\times3.9$  & 0.043 & Sbc&&Sy1.5\\
NGC4151 & 12h10m32.6s & $+$39d24m21s & 0.003319   & $6.3\times4.5$  & 0.092 & Sab&&Sy1.5\\
NGC4593 & 12h39m39.4s & $-$05d20m39s & 0.009000   & $3.9\times2.9$  & 0.082 & SBb&&Sy1\\
NGC5548 & 14h17m59.5s & $+$25d08m12s & 0.017175   & $1.4\times1.3$  & 0.068 & S0/a&&Sy1.5\\
Mrk110  & 09h25m12.9s & $+$52d17m11s & 0.035291   & $0.7\times0.4$  & 0.043 & Sc&&Sy1\\
Mrk590  & 02h14m33.5s & $-$00d46m00s & 0.026385   & $1.1\times1.0$  & 0.124 & Sa&&Sy1.2\\
Mrk817  & 14h36m22.1s & $+$58d47m39s & 0.031455   & $0.6\times0.6$  & 0.022 & SBc&&Sy1.5\\
3C120   & 04h33m11.1s & $+$05d21m16s & 0.033010   & $0.8\times0.6$  & 0.986 & S0/BLRG&&Sy1\\
PG0844$+$349 & 08h47m42.4s & $+$34d45m04s & 0.064000  & \nodata & 0.122 & Sa&&Sy1\\
\enddata
\tablenotetext{a}{
The data are taken from NED
except for the morphology of Mrk 110 and PG0844 (Bentz et al. 2009)}
\end{deluxetable}

\clearpage

\begin{deluxetable}{lccccccccc}
\tabletypesize{\footnotesize}
\tablewidth{0pt}
\tablecaption{Observational Parameters Of Target AGNs by MAGNUM telescope\label{objobsparm}}
\tablehead{
\colhead{Object} & \colhead{Filter}& \colhead{Observed period} & \multicolumn{3}{c}{Sampling interval\tablenotemark{a}}& &\multicolumn{3}{c}{Number of observations} \\
\cline{4-6}\cline{8-10}\\
 & & &\colhead{$B$}&\colhead{$V$}&\colhead{$I$}& &\colhead{$B$}&\colhead{$V$}&\colhead{$I$}
}

\startdata
NGC3227 & $BVIJ$ & 2001.11.13 $-$ 2007.07.01 &11&6&27&&62&123&17\\
NGC3516 & $BV$  & 2005.01.16 $-$ 2007.06.26 &22&14&\nodata&&17&35&\nodata\\
NGC4051 & $BVIJ$ & 2001.03.14 $-$ 2007.07.30 &10&7&38&&67&158&13\\
NGC4151 & $BVIJ$ & 2001.03.14 $-$ 2007.08.07 &13&5&36&&78&219&22\\
NGC4593 & $BV$  & 2005.01.21 $-$ 2007.07.09 &20&13&\nodata&&22&36&\nodata\\
NGC5548 & $BVIJ$ & 2001.03.23 $-$ 2007.08.20 &10&3&46&&109&302&25\\
Mrk110  & $BVIJ$ & 2003.02.06 $-$ 2007.06.11 &27&14&134&&24&56&7\\
Mrk590  & $BVIJ$ & 2001.09.11 $-$ 2007.08.08 &8&6&32&&58&128&26\\
Mrk817  & $BVIJ$ & 2003.09.04 $-$ 2007.08.18 &44&13&122&&18&64&9\\
3C120   & $BVIJ$  & 2003.02.09 $-$ 2007.08.09&26&11&34&&29&62&14\\
PG0844$+$349 & $BV$ & 2003.02.23 $-$ 2007.05.20 &135&32&\nodata&&9&22&\nodata\\
\enddata
\tablenotetext{a}{Median value of sampling interval (days)}
\end{deluxetable}

\clearpage

\begin{deluxetable}{lccc}
\tabletypesize{\footnotesize}
\tablecaption{HST/ACS/HRC Archival Images Used\tablenotemark{a}\label{BPMdata}}
\tablewidth{0pt}
\tablehead{
\colhead{Object} & \colhead{Exposure time\tablenotemark{b}} & \colhead{Proposal ID\tablenotemark{c}}\\
                 & \colhead{(sec)}                           & 
}
\startdata
NGC3227 & 300 & 9851 \\
NGC3516 & 300 & 10516 \\
NGC4051 & 300 & 9851 \\
NGC4151 & 300 & 9851 \\
NGC4593 & 300 & 10516 \\
NGC5548 & 600 & 9851 \\
Mrk110  & 600 & 9851 \\
Mrk590  & 600 & 9851 \\
Mrk817  & 300 & 9851 \\
3C120   & 300 & 9851 \\
PG0844$+$349 & 120 & 9851 \\
\enddata
\tablenotetext{a}{All images were obtained through the F550M filter.}
\tablenotetext{b}{Adopted exposure time of archival HST/ACS data}
\tablenotetext{c}{Proposal ID of archival HST/ACS data}
\end{deluxetable}

\clearpage

\begin{deluxetable}{lccccccc}
\tabletypesize{\footnotesize}
\tablewidth{0pt}
\tablecolumns{8} 
\tablecaption{Parameter Values Fitted To HST/ACS/HRC Images\label{galfitparmresult}}
\tablehead{
 & \multicolumn{2}{c}{de Vaucouleurs'} & & \multicolumn{2}{c}{Exponential} & & \\
\cline{2-3}\cline{5-6}\\
\colhead{Object}& \colhead{Scale length} & \colhead{Axis ratio} &  &\colhead{Scale length} & \colhead{Axis ratio} & \colhead{Sky level}&\colhead{Reduced $\chi^2$}
}
\startdata
NGC3227 &2.40&0.58& &54.0&0.68&2.72&1.774\\
NGC3516 &19.1&0.73& &1.44&0.61&3.89&1.318\\
NGC4051 &3.18&0.81& &43.0&0.74&3.73&1.478\\
NGC4151 &10.2&0.86& &56.6&0.80&2.30&1.764\\
NGC4593 &17.2&0.58& &2.22&0.75&3.46&1.347\\
NGC5548 &5.65&0.91& &6.52&0.81&3.48&1.184\\
Mrk110 &6.23&0.91& &\nodata&\nodata&2.58&1.143\\
Mrk590 &11.6&0.95& &\nodata&\nodata&5.46&1.241\\
Mrk817 &0.51&0.67& &3.90&0.75&4.65&1.119\\
3C120 &0.079&0.32& &1.31&0.92&5.21&1.113\\
PG0844+349&7.7&0.678& &\nodata&\nodata&0.25&1.190\\
\enddata
\tablecomments{Scale length in units of arcsec, and 
sky level in units of counts.}
\end{deluxetable}

\clearpage

\begin{deluxetable}{lcccccccc}
\tabletypesize{\footnotesize}
\tablecolumns{9}
\tablecaption{Host Galaxy Flux In The $\phi = 8''.3$ Aperture\label{MAGNUMhostresult}}
\tablewidth{0pt}
\tablehead{
 & \multicolumn{2}{c}{$B$} & & \multicolumn{2}{c}{$V$} & & \multicolumn{2}{c}{$I$}\\
\cline{2-3} \cline{5-6} \cline{8-9} \\
\colhead{Object}&\colhead{Flux}&\colhead{$N$}& &\colhead{Flux}&\colhead{$N$}& &\colhead{Flux}&\colhead{$N$}
}
\startdata
NGC3227 &4.18 $\pm$ 0.32&7& &\phn8.60 $\pm$ 0.32&7& &20.83 $\pm$ 0.54&7\\
NGC3516 &8.63 $\pm$ 0.19&3& &16.07 $\pm$ 0.28&5& &\nodata&\nodata\\
NGC4051 &4.12 $\pm$ 0.28&5& &\phn8.05 $\pm$ 0.38&7& &16.84 $\pm$ 0.78&5\\
NGC4151 &9.31 $\pm$ 0.37&7& &18.69 $\pm$ 0.90&7& &30.05 $\pm$ 0.80&8\\
NGC4593 &3.53 $\pm$ 0.07&4& &\phn7.26 $\pm$ 0.094&5& &\nodata&\nodata\\
NGC5548 &2.07 $\pm$ 0.08&6& &\phn4.23 $\pm$ 0.04&6& &\phn8.57 $\pm$ 0.14&7\\
Mrk110 &0.38 $\pm$ 0.07&7& &\phn0.735 $\pm$ 0.087&6& &\phn1.57 $\pm$ 0.11&3\\
Mrk590 &1.99 $\pm$ 0.13&8& &\phn4.75 $\pm$ 0.07&9& &10.26 $\pm$ 0.09&7\\
Mrk817 &0.46 $\pm$ 0.03&3& &\phn1.16 $\pm$ 0.07&8& &\phn2.59 $\pm$ 0.29&5\\
3C120 &1.89 $\pm$ 0.32&7& &\phn3.50 $\pm$ 0.73&8& &\phn5.93 $\pm$ 1.11&3\\
PG0844$+$349&0.42 $\pm$ 0.02&2& &\phn0.69 $\pm$ 0.16&7& &\nodata&\nodata\\
\enddata
\tablecomments{Flux in units of mJy.  $N$ is the number of images used for fitting.}
\end{deluxetable}

\clearpage

\begin{deluxetable}{lccc}
\tabletypesize{\footnotesize}
\tablecaption{Host Galaxy Color\label{hostcolor}}
\tablewidth{0pt}
\tablehead{
\colhead{Object} & \colhead{$B-V$} & \colhead{$V-I$} & \colhead{Hubble type}\\
 & \colhead{(mag)} & \colhead{(mag)} &
}
\startdata
NGC3227 &0.90 $\pm$  0.09 &   1.37 $\pm$  0.05 & SAB\\
NGC3516 &0.80 $\pm$  0.03 & \nodata & SB0\\
NGC4051 &0.85 $\pm$  0.09 &  1.21 $\pm$  0.07 & Sbc\\
NGC4151 & 0.88 $\pm$  0.07 &   0.92 $\pm$  0.06 & Sab\\
NGC4593 &0.90 $\pm$  0.03 & \nodata & SBb\\
NGC5548 & 0.89 $\pm$  0.04 &   1.17 $\pm$  0.02 & S0/a\\
Mrk110 &0.84 $\pm$  0.24 &   1.23 $\pm$  0.14 & Sc\\
Mrk590 &1.06 $\pm$  0.07 &   1.24 $\pm$  0.02 & Sa\\
Mrk817 &1.12 $\pm$  0.10 &   1.28 $\pm$  0.13 & SBc\\
3C120 & 0.79 $\pm$  0.30 &   0.98 $\pm$  0.27 & S0\\
PG0844+349 & 0.65 $\pm$  0.26 & \nodata & Sa\\
\enddata
\tablecomments{Typical color of bulge is $B-V=1.02$ (Kinney et al. 1996), 
and typical color of galaxies is $(B-V,V-I)=(0.96,1.31)$ for E, $(0.78,1.21)$ for Sab, 
and $(0.50,1.07)$ for Scd (Fukugita et al. 1995).}
\end{deluxetable}

\clearpage

\begin{deluxetable}{lccccl}
\tabletypesize{\footnotesize}
\tablecaption{Narrow Line Flux\label{NLfluxdata}}
\tablewidth{0pt}
\tablehead{
\colhead{Object}& \colhead{$[$\ion{O}{3}$]\lambda5007$} & \colhead{$[$\ion{O}{3}$]\lambda4959$} & \colhead{H$\beta$} & \colhead{H$\gamma$} &\colhead{Reference}
}
\startdata
NGC3227 &693&231&61&28& 1, 2 \\
NGC3516 &557&186&56&26& 1 \\
NGC4051 &408&136&41&18& 1\\
NGC4151 &12800&4320&1097&471& 1, 2 \\
NGC4593 &188&63&19&9& 1 \\
NGC5548 &623&208&66&30& 1, 2 \\
Mrk110 &230&77&16&7& 1, 2 \\
Mrk590 &60.4&20.1&13.9&6.3& 1, 2 \\
Mrk817 &133&44&8&4& 1, 2 \\
3C120 &819&273&86&42& 3 \\
PG0844+349 &60.3&20.1&6.1&2.8& 4 \\
\enddata
\tablerefs{
(1) Whittle et al. (1992); (2) Kaspi et al. (2005); (3) Tadhunter et al. (1993); (4) Miller et al. (1992).
}
\tablecomments{Flux in units of $10^{-15}$ergs s$^{-1}$ cm$^{-2}$. 
All the data tabulated here are corrected for the Galactic extinction. 
The $[$\ion{O}{3}$]\lambda4959$ flux and H$\gamma$ flux 
are assumed to be specific fractions of the $[$\ion{O}{3}$]\lambda5007$ flux 
and H$\beta$ flux, respectively.}
\end{deluxetable}

\clearpage

\begin{deluxetable}{lcccccc}
\tabletypesize{\footnotesize}
\tablecaption{Flux Contribution Of Narrow Lines In Each Filter\label{NLfinalresult}}
\tablewidth{0pt}
\tablehead{
\colhead{Object} & \colhead{Filter} & \colhead{$[$\ion{O}{3}$]\lambda$5007} & \colhead{$[$\ion{O}{3}$]\lambda$4959} & \colhead{H$\beta$} & \colhead{H$\gamma$} & \colhead{Total flux}
}
\startdata
NGC3227 &$B$&0.150 &  0.062 &  0.02 &  0.019 &  0.256  \\
        &$V$&0.502 &  0.077 &  0.000 &  0.000 &  0.579 \\
NGC3516 &$B$&0.100 &  0.046 &  0.020 &  0.018 &  0.184\\
        &$V$&0.474 &  0.104 &  0.000 &  0.000 &  0.577\\
NGC4051 &$B$&0.092 &  0.038 &  0.017 &  0.013 &  0.159 \\
        &$V$&0.277 &  0.035 &  0.000 &  0.000 &  0.312 \\
NGC4151 &$B$& 2.80 &  1.18 &  0.44 &  0.32 &  4.74 \\
        &$V$&9.10 &  1.31 &  0.00 &  0.00 & 10.41\\
NGC4593 &$B$&0.033 &  0.015 &  0.007 &  0.006 &  0.061\\
        &$V$&0.160 &  0.035 &  0.000 &  0.000 &  0.196\\
NGC5548 &$B$&0.081 &  0.039 &  0.020 &  0.022 &  0.162\\
        &$V$&0.606 &  0.170 &  0.005 &  0.002 &  0.783 \\
Mrk110 &$B$& 0.014 &  0.007 &  0.003 &  0.005 &  0.030\\
       &$V$& 0.248 &  0.079 &  0.012 &  0.004 &  0.343\\
Mrk590 &$B$&0.006 &  0.003 &  0.004 &  0.005 &  0.016\\
       &$V$&0.063 &  0.019 &  0.006 &  0.002 &  0.090\\
Mrk817 &$B$&0.010 &  0.005 &  0.002 &  0.003 &  0.019 \\
       &$V$&0.141 &  0.045 &  0.005 &  0.002 &  0.193 \\
3C120 &$B$&0.056 &  0.028 &  0.019 &  0.031 &  0.135 \\
      &$V$&0.874 &  0.279 &  0.059 &  0.023 &  1.235 \\
PG0844+349 &$B$& 0.001 &  0.000 &  0.000 &  0.002 &  0.004\\
           &$V$&0.067 &  0.022 &  0.007 &  0.000 &  0.096 \\
\enddata
\tablecomments{Flux in units of mJy.}
\end{deluxetable}

\clearpage

\begin{deluxetable}{lcccccc}
\tabletypesize{\footnotesize}
\tablecaption{Statistics Of Linear Flux To Flux Fit\label{linefittable}}
\tablewidth{0pt}
\tablehead{
\colhead{Object} & \colhead{Paired bands} & \colhead{$a$\tablenotemark{a}} & \colhead{$b$\tablenotemark{a}} & \colhead{$N$} & \colhead{Reduced $\chi^2$} & \colhead{$R_V$($\times100$)}
}
\startdata
NGC3227 &$BV$& $0.96 \pm 0.02$ &\phn$-5.26\pm 0.22$&54& 0.88 & 1.70 \\
        &$VI$& $0.78 \pm 0.03$&\phn$-7.58 \pm 0.82$&17&3.30 & 4.76\\
NGC3516 &$BV$&$1.02 \pm 0.02$&\phn$-8.37 \pm 0.44$&9&0.87 & 1.43\\
NGC4051 &$BV$& $0.99 \pm 0.02$&\phn$-4.08\pm 0.23$&55&1.52 & 1.85\\
        &$VI$& $0.96 \pm 0.04$&\phn$-8.46 \pm 0.79$&9&1.90 & 4.94\\
NGC4151 &$BV$& $1.03 \pm 0.01$&$-16.0 \pm 0.1$&36&8.47 & 0.63\\
        &$VI$&$0.79 \pm 0.01$&\phn\phs$3.16 \pm 0.25$&19&28.53 & 2.03\\
NGC4593 &$BV$&$1.05 \pm 0.01$&\phn$-4.05 \pm 0.14$&16&1.92 & 1.67\\
NGC5548 &$BV$&$1.01 \pm 0.01$&\phn$-2.60 \pm 0.04$&81&1.65 & 0.88\\
        &$VI$& $0.70 \pm 0.02$ &\phn$-1.14 \pm 0.22$&20&5.74 & 5.47\\
Mrk110 &$BV$&$0.94 \pm 0.01$&\phn$-0.60 \pm 0.04$ &11& 3.58 & 0.97\\
       &$VI$& $0.92 \pm 0.02$ & \phn$-0.53 \pm 0.10$ &7&2.91 & 4.71\\
Mrk590 &$BV$&$0.97 \pm 0.06$&\phn$-2.85 \pm 0.32$&19&1.05 & 5.51\\
       &$VI$&$0.77 \pm 0.07$&\phn$-3.08 \pm 0.73$&17&4.27 & 10.70\\
Mrk817 &$BV$&$0.93 \pm 0.03$&\phn$-0.66 \pm 0.19$&8&0.87 & 1.44\\
       &$VI$& $1.07 \pm 0.05$&\phn$-2.53 \pm 0.42$&6&0.80 & 2.43\\
3C120 &$BV$&$1.11 \pm 0.02$&\phn$-3.40 \pm 0.26$ &25&1.70 & 1.27\\
      &$VI$& $1.04 \pm 0.01$&\phn$-2.49\pm 0.23$ &11& 8.05 & 1.98\\
PG0844+349 &$BV$& $0.87 \pm 0.03$ &\phn $-0.38 \pm 0.21$ &8& 1.90 & 2.11\\
\enddata
\tablenotetext{a}{$f_Y=a\times f_X+b$. ($\lambda_Y < \lambda_X$)}
\tablecomments{$N$ is the number of data used for fitting. $R_V$ is the ratio of 
fitting residual to flux variation defined as 
$R_V=\sigma _V^{\rm res} /\Delta V$ 
where $\sigma _V^{\rm res}$ is the standard deviation of the residuals
from the best-fit regression line, and $\Delta V=f_V^{\rm max}-f_V^{\rm min}$ is 
the amplitude of flux variation.}
\end{deluxetable}

\clearpage

\begin{deluxetable}{lcccccccc}
\tabletypesize{\footnotesize}
\tablecaption{Linear $V$ Flux to $I$ Flux Fit With and Without Dust Correction\label{Idustcontritable}}
\tablewidth{0pt}
\tablehead{
& & \multicolumn{4}{c}{With correction} & & \multicolumn{2}{c}{Without correction} \\
\cline{3-6}\cline{8-9}\\
\colhead{Object}  &\colhead{$f_J^{HN}$\tablenotemark{a}} & \colhead{$T_d$\tablenotemark{b}} & \colhead{$\alpha_{VJ}$} & \colhead{$\alpha_{VI}$}&\colhead{Reduced $\chi^2$} &  & \colhead{$\alpha_{VI}$} & \colhead{Reduced $\chi^2$}    
}
\startdata
NGC3227 &47.9&1700&$-0.6$&$-0.64$&1.52&&$-0.67$&3.30 \\
NGC4051 &37.8&1950&\phs$0.0$&$-0.08$& 0.64&&$-0.11$ & 1.90   \\
NGC4151 &51.9&1650&$-0.6$&$-0.66$&4.48&&$-0.65$&28.5  \\
NGC5548 &16.4&1700&$-0.2$&$-0.57$&1.34&&$-0.97$&5.74 \\
Mrk110  &1.96&2050&$-0.6$&$-0.42$&1.35&&$-0.23$&2.91 \\
Mrk817  &5.73&1850&$-0.7$&\phs$0.13$ &0.15&&\phs$0.19$& 0.80\\
3C120   &7.30&1850&$-0.3$&$-0.24$&1.79&&\phs$0.10$&8.05\\
\enddata
\tablenotetext{a}{Flux of HOST$+$NL component in the $J$ band in units of mJy.}
\tablenotetext{b}{Dust temperature in units of $K$.}
\end{deluxetable}

\clearpage

\begin{deluxetable}{llll}
\tabletypesize{\footnotesize}
\tablecaption{Flux Contribution of the Host $+$ Narrow-Line Component\label{hostcont}}
\tablewidth{0pt}
\tablehead{
 & \multicolumn{3}{c}{$f^{\rm HOST+NL}/f^{\rm ave}$}\\
\cline{2-4}\\
\colhead{Object}
 & \multicolumn{1}{c}{$B$}
 & \multicolumn{1}{c}{$V$}
 & \multicolumn{1}{c}{$I$}
}
\startdata
     NGC3227 & 0.495 $\pm$ 0.037 & 0.622 $\pm$ 0.023 & 0.726 $\pm$ 0.019\\
     NGC3516 & 0.564 $\pm$ 0.012 & 0.714 $\pm$ 0.014 & \multicolumn{1}{c}{\nodata} \\ 
     NGC4051 & 0.482 $\pm$ 0.032 & 0.643 $\pm$ 0.030 & 0.745 $\pm$ 0.034\\
     NGC4151 & 0.437 $\pm$ 0.017 & 0.668 $\pm$ 0.023 & 0.649 $\pm$ 0.017\\ 
     NGC4593 & 0.472 $\pm$ 0.010 & 0.681 $\pm$ 0.009 & \multicolumn{1}{c}{\nodata} \\ 
     NGC5548 & 0.565 $\pm$ 0.021 & 0.789 $\pm$ 0.025 & 0.837 $\pm$ 0.013\\
      Mrk110 & 0.137 $\pm$ 0.024 & 0.287 $\pm$ 0.029 & 0.297 $\pm$ 0.021\\ 
      Mrk590 & 0.878 $\pm$ 0.056 & 0.920 $\pm$ 0.014 & 0.951 $\pm$ 0.008\\
      Mrk817 & 0.091 $\pm$ 0.006 & 0.209 $\pm$ 0.013 & 0.322 $\pm$ 0.036\\
       3C120 & 0.183 $\pm$ 0.029 & 0.368 $\pm$ 0.060 & 0.393 $\pm$ 0.073\\
PG0844$+$349 & 0.077 $\pm$ 0.004 & 0.118 $\pm$ 0.024 & \multicolumn{1}{c}{\nodata} \\
\enddata
%\tablenotetext{a}{}
\tablecomments{
$f^{\rm HOST+NL}/f^{\rm ave}$ is 
the contribution of the host-galaxy plus narrow-line component
to the average flux of the light curve data. 
The error comes from that of the HOST$+$NL flux.
}
\end{deluxetable}

\clearpage

\begin{deluxetable}{lcccc}
\tabletypesize{\footnotesize}
\tablecaption{Color and Spectral Index Of AGN Continuum\label{variablecolor}}
\tablewidth{0pt}
\tablehead{
\colhead{Object} & \colhead{$B-V$} & \colhead{$\alpha_{BV}$} & \colhead{$V-I$} & \colhead{$\alpha_{VI}$} \\
\colhead{} & \colhead{(mag)} &  &\colhead{(mag)} &
}
\startdata
NGC3227 &$0.16 \pm  0.02$&$-0.17 \pm 0.07$&$0.66 \pm 0.04$&$-0.64 \pm 0.11$\\
NGC3516 &$0.10 \pm  0.02$ &\phs$ 0.08 \pm 0.08 $&\nodata&\nodata \\
NGC4051 &$0.13 \pm  0.02$&$-0.05 \pm 0.11$&$0.37 \pm 0.05$&$-0.08 \pm 0.12$\\
NGC4151 &$0.09 \pm  0.01$&\phs$0.13 \pm 0.03$&$0.67 \pm 0.01$&$-0.66 \pm 0.02$\\
NGC4593 &$0.06 \pm  0.02$&\phs$0.24 \pm 0.08 $&\nodata&\nodata \\
NGC5548 &$0.11 \pm  0.01$&\phs$0.04 \pm 0.04$&$0.63 \pm 0.04$&$-0.57 \pm 0.10$\\
Mrk110  &$0.19 \pm  0.02$&$-0.32 \pm 0.09$&$0.57 \pm 0.02$&$-0.41 \pm 0.06$\\
Mrk590  &$0.15 \pm  0.07$&$-0.12 \pm 0.31$&$0.69 \pm 0.20$&$-0.72 \pm 0.50$\\
Mrk817  &$0.19 \pm  0.03$&$-0.32 \pm 0.13$&$0.35 \pm 0.05$&\phs$0.13 \pm 0.14$ \\
3C120   &$0.02 \pm  0.02$&\phs$0.45 \pm 0.08 $&$0.50 \pm 0.02$&$-0.24 \pm 0.04$\\
PG0844+349 &$0.27 \pm  0.05$& $-0.65 \pm 0.24$ &\nodata&\nodata \\
\enddata
\tablecomments{
Color and spectral index are derived from equations \ref{colorXY} 
and \ref{alphaXY}, respectively, in the text.
}
\end{deluxetable}

\clearpage

\begin{deluxetable}{lccc}
\tabletypesize{\footnotesize}
\tablecaption{Comparison Of Internal Extinction\tablenotemark{a}\label{internalextinctioncomparison}}
\tablewidth{0pt}
\tablehead{
\colhead{Object} & \colhead{Our estimate} & \colhead{Literature} & \colhead{Reference}
}
\startdata
NGC3227 &$0.16 \pm 0.02$&$0.18$ & 1\\ 
NGC4151 &$0.090 \pm 0.008$&$0.091 \pm 0.007$ & 2\\
NGC5548 &$0.109 \pm 0.008$&$0.073 \pm 0.018$ & 2\\
3C120 &$0.017 \pm 0.018$&$0.417 \pm 0.630$ & 2\\
\enddata
\tablerefs{
(1) Crenshaw et al. (2001); (2) Paltani et al. (1996).
}
\tablenotetext{a}{Extinction is the value of $E_{B-V}$.}
\end{deluxetable}


\clearpage

\begin{thebibliography}{}
\bibitem[Bahcall et al. (1997)]{1997ApJ...479..642B}Bahcall, J. N.,
Kirhakos, S., Saxe, D. H., \& Schneider, D. P. 1997, \apj, 479, 642
\bibitem[Bentz et al. (2006)]{2006ApJ...644..133B} Bentz, M. C., Peterson, 
B. M., Pogge, R. W., Vestergaard, M., \& Onken, C. A.\ 2006, \apj, 644, 133 
\bibitem[Bentz et al. (2009)]{2009ApJ...697..160B} Bentz, M. C., Peterson, 
B. M., Netzer, H., Pogge, R. W., \& Vestergaard, M.\ 2009, \apj, 697, 160
\bibitem[Bessell(1979)]{1979PASP...91..589B} Bessell, M. S.\ 1979, PASP, 
91, 589 
\bibitem[Bregman(1990)]{1990A&ARv...2..125B} Bregman, J. N.\ 1990, \aapr, 2, 125 
\bibitem[Choloniewski(1981)]{1981AcA....31..293C} Choloniewski, J.\ 1981, 
Acta Astronomica, 31, 293 
\bibitem[Courvoisier et 
al.(1996)]{1996A&A...308L..17C} Courvoisier, T. J.-L., Paltani, S., \& Walter, R.\ 1996, \aap, 308, L17 
\bibitem[Crenshaw et al. (1996)]{1996ApJ...470..322C} Crenshaw, D. M., et 
al.\ 1996, \apj, 470, 322 
\bibitem[Crenshaw et al.(2001)]{2001ApJ...555..633C} Crenshaw, D. M., 
Kraemer, S. B., Bruhweiler, F. C., \& Ruiz, J. R.\ 2001, \apj, 555, 633
\bibitem[de 
Jong(1996)]{1996A&A...313..377D} de Jong, R.~S.\ 1996, \aap, 313, 377 
\bibitem[Doroshenko \& Lyutyi (1994)]{1994AstL...20..606D}Doroshenko, V. T.,
 \& Lyutyi, V. M. 1994, Astron. Lett., 20, 606
\bibitem[Draine  \& Lee(1984)]{1984ApJ...285...89D} Draine, B. T., \& Lee, H. M.\ 1984, \apj, 285, 89
\bibitem[Ferruit et al.(1998)]{1998ApJ...509..646F} Ferruit, P., Wilson, 
A.~S., \& Mulchaey, J.~S.\ 1998, \apj, 509, 646 
\bibitem[Fukugita et al.(1995)]{1995PASP..107..945F} Fukugita, M., 
Shimasaku, K., \& Ichikawa, T.\ 1995, PASP, 107, 945 
\bibitem[Giveon et al. (1999)]{1999MNRAS.306..637G} Giveon, U., Maoz, D., 
Kaspi, S., Netzer, H., \& Smith, P. S.\ 1999, \mnras, 306, 637 
\bibitem[Hawkins(1993)]{1993Natur.366..242H} Hawkins, M. R. S.\ 1993, \nat, 
366, 242 
\bibitem[Hunt et al.(1998)]{1998AJ....115.2594H} Hunt, L. K., Mannucci, F., 
Testi, L., Migliorini, S., Stanga, R. M., Baffa, C., Lisi, F., 
\& Vanzi, L.\ 1998, \aj, 115, 2594
\bibitem[Jahnke et al.(2004)]{2004MNRAS.352..399J} Jahnke, K., Kuhlbrodt, 
B., \& Wisotzki, L.\ 2004, \mnras, 352, 399 
\bibitem[Jefferys(1980)]{1980AJ.....85..177J} Jefferys, W. H.\ 1980, \aj, 
85, 177 
\bibitem[Jefferys(1981)]{1981AJ.....86..149J} Jefferys, W. H.\ 1981, \aj, 
86, 149 
\bibitem[Kaspi et al. (1996)]{1996ApJ...470..336K} Kaspi, S., et al.\ 1996, 
\apj, 470, 336 
\bibitem[Kaspi et al. (2005)]{2005ApJ...629...61K} Kaspi, S., Maoz, D., 
Netzer, H., Peterson, B. M., Vestergaard, M., \& Jannuzi, B. T.\ 2005, 
\apj, 629, 61 
\bibitem[Kawaguchi et al.(1998)]{1998ApJ...504..671K} Kawaguchi, T., 
Mineshige, S., Umemura, M., \& Turner, E. L.\ 1998, \apj, 504, 671 
\bibitem[Kinney et al.(1996)]{1996ApJ...467...38K} Kinney, A. L., Calzetti, 
D., Bohlin, R. C., McQuade, K., Storchi-Bergmann, T., \& Schmitt, H. R.\
1996, \apj, 467, 38 
\bibitem[Kobayashi et al.(1993)]{1993ApJ...404...94K} Kobayashi, Y., Sato, 
S., Yamashita, T., Shiba, H., \& Takami, H.\ 1993, \apj, 404, 94 
\bibitem[Kobayashi et al.(1998)]{1998SPIE.3352..120K} Kobayashi, Y., et 
al.\ 1998, \procspie, 3352, 120 
\bibitem[Kobayashi et al.(1998)]{1998SPIE.3354..769K} Kobayashi, Y., 
Yoshii, Y., Peterson, B. A., Minezaki, T., Enya, K., Suganuma, M., 
\& Yamamuro, T.\ 1998, \procspie, 3354, 769 
\bibitem[Kobayashi et al.(2003)]{2003SPIE.4837..954K} Kobayashi, Y., et 
al.\ 2003, \procspie, 4837, 954 
\bibitem[Korista et al. (1995)]{1995ApJS...97..285K} Korista, K. T., et al.\ 
1995, \apjs, 97, 285 
\bibitem[Kormendy et al.(2006)]{2006ApJ...642..765K} Kormendy, J., Cornell, 
M.~E., Block, D.~L., Knapen, J.~H., \& Allard, E.~L.\ 2006, \apj, 642, 765 
\bibitem[Kraemer et al. (1999)]{1999ApJ...520..564K} Kraemer, S. B., Ho, 
L. C., Crenshaw, D. M., Shields, J. C., \& Filippenko, A. V.\ 1999, \apj, 
520, 564 
\bibitem[Kriss \& Krolik(1994)]{1994ASPC...54..111K} Kriss, G. A., \& 
Krolik, J. H.\ 1994, The Physics of Active Galaxies, 54, 111 
\bibitem[Krist(1995)]{1995ASPC...77..349K} Krist, J.\ 1995, Astronomical 
Data Analysis Software and Systems IV, 77, 349 
\bibitem[Krolik et al.(1991)]{1991ApJ...371..541K} Krolik, J. H., Horne, 
K., Kallman, T. R., Malkan, M. A., Edelson, R. A., 
\& Kriss, G. A.\ 1991, \apj, 371, 541 
\bibitem[Lodders 
\& Fegley(1999)]{1999IAUS..191..279L} Lodders, K., \& Fegley, B., Jr.\ 1999, Asymptotic Giant Branch Stars, 191, 279 
\bibitem[Malkan 
\& Sargent(1982)]{1982ApJ...254...22M} Malkan, M. A., \& Sargent, W. L. W.\ 1982, \apj, 254, 22 
\bibitem[Minezaki et al.(2004)]{2004ApJ...600L..35M} Minezaki, T., Yoshii, 
Y., Kobayashi, Y., Enya, K., Suganuma, M., Tomita, H., Aoki, T., \& 
Peterson, B. A.\ 2004, \apjl, 600, L35 
\bibitem[Paltani \& Walter(1996)]{1996A&A...312...55P} Paltani, S., \& 
Walter, R.\ 1996, \aap, 312, 55 
\bibitem[Peng et al. (2002)]{2002AJ....124..266P} Peng, C. Y., Ho, L. C., 
Impey, C. D., \& Rix, H.-W.\ 2002, \aj, 124, 266 
\bibitem[Pereyra et al. (2006)]{2006ApJ...642...87P}Pereyra, N. A.,
Vanden Berk, D. E., Turnshek, D. A., Hillier, D. J.,
Wilhite, B. C., Kron, R. G., Schneider, D. P., \& Brinkmann, J.
2006, \apj, 642, 87
\bibitem[Peterson \& Wandel(2000)]{2000ApJ...540L..13P} Peterson, B. M., \& 
Wandel, A.\ 2000, \apjl, 540, L13 
\bibitem[Peterson et al.(2004)]{2004ApJ...613..682P} Peterson, B. M., et 
al.\ 2004, \apj, 613, 682 
\bibitem[Rees(1984)]{1984ARA&A..22..471R} Rees, M. J.\ 1984, \araa, 22, 471
\bibitem[Reichert et al. (1994)]{1994ApJ...425..582R} Reichert, G. A., et 
al.\ 1994, \apj, 425, 582 
\bibitem[Rodriguez-Pascual et al. (1997)]{1997ApJS..110....9R} 
Rodriguez-Pascual, P. M., et al.\ 1997, \apjs, 110, 9 
\bibitem[Romano \& Peterson(1998)]{1998astro.ph..6190R} Romano, P., \& 
Peterson, B. M.\ 1998, ArXiv Astrophysics e-prints, arXiv:astro-ph/9806190 
\bibitem[Salpeter (1977)]{1977ARA&A..15..267S} Salpeter, E. E.
 1977, \araa, 15, 267
\bibitem[Santos-Lleo et al. (1995)]{1995MNRAS.274....1S} Santos-Lleo, M., 
Clavel, J., Barr, P., Glass, I. S., Pelat, D., Peterson, B. M., \& 
Reichert, G.\ 1995, \mnras, 274, 1
\bibitem[Shakura \& Syunyaev(1973)]{1973A&A....24..337S} Shakura, N. I., \& 
Syunyaev, R. A.\ 1973, \aap, 24, 337 
\bibitem[Schlegel et al.(1998)]{1998ApJ...500..525S} Schlegel, D. J., 
Finkbeiner, D. P., \& Davis, M.\ 1998, \apj, 500, 525 
\bibitem[Sch{\"o}del et al.(2003)]{2003ApJ...596.1015S} Sch{\"o}del, R., 
Ott, T., Genzel, R., Eckart, A., Mouawad, N., \& Alexander, T.\ 2003, \apj, 
596, 1015 
\bibitem[Schmitt et al.(2003)]{2003ApJ...597..768S} Schmitt, H. R., Donley, 
J. L., Antonucci, R. R. J., Hutchings, J. B., Kinney, A. L., 
\& Pringle, J. E.\ 2003, \apj, 597, 768
\bibitem[Shields(1978)]{1978Natur.272..706S} Shields, G. A.\ 1978, \nat, 
272, 706
\bibitem[Smith 
\& Hoffleit(1963)]{1963AJ.....68S.292S} Smith, H. J., \& Hoffleit, D.\ 1963, \aj, 68, 292 
\bibitem[Storey \& Zeippen(2000)]{2000MNRAS.312..813S} Storey, P. J., \& 
Zeippen, C. J.\ 2000, \mnras, 312, 813 
\bibitem[Suganuma et al. (2006)]{2006ApJ...639...46S} Suganuma, M., et al.\ 
2006, \apj, 639, 46 
\bibitem[Tadhunter et al. (1993)]{1993MNRAS.263..999T} Tadhunter, C. N., 
Morganti, R., di Serego-Alighieri, S., Fosbury, R. A. E., \& Danziger, 
I. J.\ 1993, \mnras, 263, 999 
\bibitem[Taylor et al. (2005)]{2005ApJ...630..784T} Taylor, V. A.,
Jansen, R. A., Windhorst, R. A., Odewahn, S. C., \& Hibbard, J. E.
\ 2005, \apj, 630, 784
\bibitem[Terlevich et al.(1992)]{1992MNRAS.255..713T} Terlevich, R., 
Tenorio-Tagle, G., Franco, J., \& Melnick, J.\ 1992, \mnras, 255, 713 
\bibitem[Tomita (2005)]{tomita05} Tomita, H.\ 2005, Ph.D. dissertation, Department of
Astronomy, Graduate School of Science, The University of Tokyo
\bibitem[Tomita et al. (2006)]{2006ApJ...652L..13T} Tomita, H., et al.\ 
2006, \apjl, 652, L13 
\bibitem[Trevese et al. (1994)]{1994ApJ...433..494T} Trevese, D., Kron, 
R. G., Majewski, S. R., Bershady, M. A., \& Koo, D. C.\ 1994, \apj, 433, 
494 
\bibitem[Tr{\`e}vese 
\& Vagnetti(2002)]{2002ApJ...564..624T} Tr{\`e}vese, D., \& Vagnetti, F.\ 2002, \apj, 564, 624 
\bibitem[Vanden Berk et al. (2001)]{2001AJ....122..549V} Vanden Berk, D. E., 
et al.\ 2001, \aj, 122, 549 
\bibitem[Vanden Berk et al. (2004)]{2004ApJ...601..692V} Vanden Berk, D. E., 
et al.\ 2004, \apj, 601, 692 
\bibitem[Veilleux \& Osterbrock(1987)]{1987ApJS...63..295V} Veilleux, S., 
\& Osterbrock, D. E.\ 1987, \apjs, 63, 295 
\bibitem[Wamsteker et al. (1990)]{1990ApJ...354..446W} Wamsteker, W., et 
al.\ 1990, \apj, 354, 446 
\bibitem[Webb \& Malkan(2000)]{2000ApJ...540..652W} Webb, W., \& Malkan, 
M.\ 2000, \apj, 540, 652 
\bibitem[Whittle(1992)]{1992ApJS...79...49W} Whittle, M.\ 1992, \apjs, 79, 
49 
\bibitem[Wilhite et al. (2005)]{2005ApJ...633..638W} Wilhite, B. C., Vanden 
Berk, D. E., Kron, R. G., Schneider, D. P., Pereyra, N., Brunner, R. J., 
Richards, G. T., \& Brinkmann, J. V.\ 2005, \apj, 633, 638 
\bibitem[Wilhite et al.(2008)]{2008MNRAS.383.1232W} Wilhite, B. C., 
Brunner, R. J., Grier, C. J., Schneider, D. P., 
\& vanden Berk, D. E.\ 2008, \mnras, 383, 1232
\bibitem[Winkler et al. (1992)]{1992MNRAS.257..659W} Winkler, H., Glass, 
I. S., van Wyk, F., Marang, F., Jones, J. H. S., Buckley, D. A. H., \& 
Sekiguchi, K.\ 1992, \mnras, 257, 659 
\bibitem[Winkler(1997)]{1997MNRAS.292..273W} Winkler, H.\ 1997, \mnras, 
292, 273
\bibitem[Wold et al.(2007)]{2007MNRAS.375..989W} Wold, M., Brotherton, 
M. S., \& Shang, Z.\ 2007, \mnras, 375, 989
\bibitem[Yoshida \& Ohtani(1993)]{1993PASJ...45..407Y} Yoshida, M., \& 
Ohtani, H.\ 1993, \pasj, 45, 407
\bibitem[Yoshii(2002)]{2002ntto.conf..235Y} Yoshii, Y.\ 2002, New Trends in 
Theoretical and Observational Cosmology, 235
\bibitem[Yoshii et al.(2003)]{2003AAS...202.3803Y} Yoshii, Y., Kobayashi, 
Y., \& Minezaki, T.\ 2003, Bulletin of the American Astronomical Society, 35, 752
\end{thebibliography}
\end{document}